# Radial-Component Predominant-Mode Inversion of Rayleigh Waves: Application to DAS-based Site Characterization

Mrinal Bhaumik[1] and Brady R. Cox
Department of Civil and Environmental Engineering, Utah State University, Logan, Utah, USA

## Abstract

Distributed Acoustic Sensing (DAS) has emerged as a transformative technology for near-surface site characterization, offering dense spatial sampling of wavefields over kilometer-long fiber-optic cables. While DAS measurements have already proven useful for extracting and inverting Rayleigh-wave dispersion data for subsurface imaging, a component-consistent approach between the experimental data and the inversion forward problem would improve inversion accuracy. When a vertical surface source is activated along the fiber, DAS measures only the in-line (radial) component of Rayleigh-wave motion. However, conventional surface-wave techniques, such as Multichannel Analysis of Surface Waves (MASW), primarily measure the vertical component of Rayleigh-wave motion. Dispersion data extracted from radial-component waveforms can differ from dispersion data extracted from vertical-component waveforms. These differences become more pronounced in complex stratigraphic conditions. Hence, when inverting radial-component dispersion data extracted from DAS waveforms a component-consistent forward problem is desired to retrieve the most accurate shear wave velocity ($V_S$) profiles. This study presents a radial-component predominant-mode (RCPM) inversion framework designed for DAS-based surface-wave analysis that explicitly accounts for source–receiver directivity and modal sensitivity of the Rayleigh-wave radial component. The proposed RCPM approach focuses on matching measured dominant-amplitude radial dispersion trends with the theoretical mode exhibiting the maximum modal amplitude. As a result, the RCPM framework eliminates the need for explicit modal indexing, provides a component-consistent interpretation of radial-component dispersion data, and substantially reduces reliance on subjective analyst-driven modal interpretations. The RCPM approach is systematically evaluated using three synthetic ground models and two field DAS datasets. The synthetic results demonstrate that modal energy distribution differs significantly between vertical and radial components in the presence of strong velocity contrasts, low-velocity layers, and high-velocity layers, and that conventional inversion approaches may misinterpret modal behavior, resulting in less accurate $V_S$ profiles. In contrast, the proposed RCPM method consistently captures the correct modal response and yields stable and reliable $V_S$ profiles across different inversion parameterizations. The RCPM inversion approach is further validated using two field DAS datasets for which the inverted $V_S$ profiles show good agreement with independent invasive measurements, including cone penetration test and borehole data.

[1] Corresponding author
Email: mrinal.bhaumik@usu.edu

# 1. Introduction

Surface-wave methods are widely used to estimate subsurface shear wave velocity ($V_s$) profiles for site characterization due to their non-invasive nature, rapid deployment, cost-effectiveness, and ability to represent the subsurface over relatively large volumes (Kramer 1996). Surface-waves in layered media are naturally dispersive, meaning that waves of different wavelengths propagate at different phase velocities. Longer wavelengths penetrate deeper into the subsurface, whereas shorter wavelengths are primarily sensitive to shallower layers. Consequently, each wavelength carries information about a specific depth range of the subsurface. Therefore, the variation of phase velocity with frequency (dispersion) provides insight into depth-dependent subsurface stratigraphy. In practice, surface-waves are recorded at discrete locations along the ground surface, and their dispersion characteristics are extracted through wavefield transformation. The resulting dispersion data are then inverted to obtain the corresponding $V_s$ profile (Nazarian et al. 1983; Park et al. 1999). Both Rayleigh and Love waves can be utilized for surface wave testing; however, Rayleigh wave-based approaches are most commonly employed in practice (Foti et al. 2018). Rayleigh waves exhibit retrograde elliptical particle motion aligning in the plane of propagation, consisting of coupled vertical and radial (in-line) displacement components (Aki and Richards 2002). Conventional surface-wave techniques, such as the active source Multichannel Analysis of Surface Waves (MASW) (Park et al. 1999; Foti 2000) and passive Microtremor Array Measurements (MAM) (Okada 2003), primarily utilize the vertical component of Rayleigh-wave motion. In MASW surveys, the vertical component is more commonly preferred because vertical geophones are easier to deploy, provide more stable coupling, and do not require strict alignment of sensors and sources. The radial component can also be measured using horizontal geophones oriented in the in-line direction of the array, combined with either vertical or in-line hammer sources. Nevertheless, conventional MASW surveys often ignore the radial component measurement due to practical limitations in data acquisition, such as the difficulty of deploying and aligning horizontal geophones and sources in field conditions. Although the vertical component remains the standard choice, the radial component can exhibit distinct and advantageous modal characteristics, including enhanced sensitivity to higher modes (Boaga et al. 2013; Vantassel and Cox 2022). Dal Moro (2015, 2020) and Dal Moro et al. (2017) presented several examples of real-field and synthetic cases highlighting clear differences in modal energy distribution between vertical and radial component dispersion images.

Recent advances in Distributed Acoustic Sensing (DAS) have introduced a transformative approach to seismic data acquisition for near-surface characterization (Daley et al. 2013; Soga and Luo 2018; Ajo-Franklin et al. 2019; Hubbard et al. 2022; Yust et al. 2024). DAS systems utilize standard fiber-optic cables interrogated by coherent laser pulses to measure dynamic strain (or strain rate) along the fiber. This effectively converts the fiber into a dense array of sensors with meter-scale or finer spatial sampling over 10's of kilometers of spatial coverage. The DAS response is inherently directional and sensitive only to motion projected along the fiber axis. Therefore, when a surface source is activated along the fiber, DAS measurements predominantly capture the radial component of Rayleigh-wave motion. Unlike conventional geophones, which record particle velocity, DAS measures the axial strain component, corresponding to the spatial gradient of displacement along the fiber direction (Hubbard et al. 2022). Studies have shown that dispersion data obtained from in-line geophones and DAS measurements are broadly consistent with one another (Vantassel et al. 2022). However, because DAS is only capable of measuring radial motions, one should not expect dispersion data extracted from DAS waveforms to always agree well with dispersion data extracted from vertical geophones.

The radial component of Rayleigh waves is not merely a rotated version of the vertical component; rather, it may exhibit fundamentally different modal behavior. Similar to the vertical component, the radial behavior of Rayleigh waves is site dependent, and the modal energy transition may exhibit differently than vertical components due to ellipticity (Boaga et al. 2013; Dal Moro et al. 2017; Vantassel and Cox 2022; Bhaumik and Naskar 2024c). Consequently, dispersion images derived from radial-component DAS data often reveal modal patterns that differ from those obtained using vertical-component measurements. These differences become more pronounced in complex stratigraphic conditions, such as the presence of strong velocity contrasts, low-velocity layers (LVLs), or high-velocity layers (HVLs). In many cases, especially due to embedded HVLs and high velocity contrasts, the dispersion spectrum may exhibit smooth transitions between modes, making mode identification and indexing challenging (O'Neill and Matsuoka 2005). Even when apparent modal transition points exist, as in some LVL scenarios, modal excitation may not follow a sequential pattern. Instead of transitioning from the fundamental mode (R0) to the first higher mode (R1), the energy may directly shift to other higher-order modes (e.g., R2 or R3), effectively skipping intermediate modes. This non-sequential behavior is a direct consequence of the underlying modal energy distribution and cannot be reliably inferred from experimental dispersion images alone. In practice, analysts often assign modal branches based on visual inspection of the dispersion data, or allow inversion algorithms to select modes solely based on minimizing dispersion misfit. However, such approaches can lead to multiple plausible solutions with similar misfit values but significantly different subsurface profiles. As a result, conventional multimodal inversion, which relies on explicit mode indexing, becomes highly sensitive to interpretation. Although DAS-based dispersion data has been widely applied in near-surface investigations, most existing studies (Luo et al. 2020; Lancelle et al. 2021; Shragge et al. 2021; Rossi et al. 2022; Jiang et al. 2023; Yust et al. 2024; Roshdy et al. 2025; Wang et al. 2025) continue to adopt inversion strategies relying on either a fundamental mode interpretation or a multi-modal interpretation with explicit modal indexing. Incorrect mode identification can lead to physically inconsistent velocity profiles, essentially undermining the true advantages of high-resolution DAS data. This highlights the need for a component-consistent inversion framework that directly accounts for the modal behavior observed in radial-component DAS data and reduces the ambiguity in manual modal interpretations. Bhaumik and Cox (2026) presented an amplitude informed predominant-mode inversion framework for vertical component Rayleigh wave data which address the ambiguity in mode indexing by considering modal energy on the surface. The same formulation, however, doesn't inherently apply to the radial component measurement.

To address these discrepancies, this study proposes a radial-component predominant-mode inversion framework designed for DAS-based surface-wave measurements that explicitly accounts for the sensitivity of the Rayleigh-wave radial component. The proposed approach eliminates the need for manual modal interpretation and instead focuses on matching dominant radial component dispersion trends with amplitude-informed modal energy distributions. The methodology is evaluated through three synthetic models and two field DAS datasets. The synthetic profiles represent common geotechnical scenarios, including monotonically increasing velocity with depth, strong velocity contrasts, and the presence of LVLs and HVLs. For each case, the dispersion behavior of both vertical and radial components is first analyzed and validated through numerical simulations. Subsequently, radial component dispersion targets, constructed by simulating multiple active and passive field measurements, are inverted using a differential evolution (DE)-based global optimization scheme. To highlight the importance of using this new radial-component predominant-mode (RCPM) approach for DAS measurements, the RCPM inversions are compared with a conventional multi-modal (CMM) approach where mode indexing must be assigned manually by the analyst. Depending upon the complexity of modal behavior, one or more plausible CMM interpretation scenarios are considered for the synthetic models to account for the subjectivity associated with manual mode indexing. The results demonstrate that for some subsurface conditions the CMM approaches can yield ambiguous interpretations due to challenges in radial component mode identification. In contrast, the proposed RCPM framework provides more consistent and physically meaningful results by incorporating radial component modal energy into the inversion process. These findings establish the importance of a component-consistent, amplitude-informed, predominant-mode inversion framework for DAS measurements. The methodology is further validated using two field DAS datasets, where the inversion results show good agreement with available geological and borehole information. By integrating DAS measurement physics into the inversion framework, the proposed methodology aims to improve the reliability and robustness of subsurface $V_s$ estimation, particularly in scenarios where higher modes are significant and conventional approaches become ambiguous.

## 2. Computation of the Radial-Component Predominant-Mode (RCPM)

This section presents the methodology adopted to compute the radial-component predominant-mode of Rayleigh waves. Herein, we define the predominant mode as the mode with the highest theoretical energy/amplitude at the ground surface for any given frequency, and note that the predominant mode can transition between modes for many subsurface conditions (Bhaumik and Cox 2026). Among several forward modeling approaches, including the transfer matrix method, stiffness matrix method, fast-delta matrix method, and thin layer method (TLM), the TLM is implemented in this study. The primary advantage of the TLM lies in its formulation of the forward problem as an eigenvalue problem, facilitating simultaneous computation of modal wavenumbers (eigenvalues) and corresponding mode shapes (eigenvectors). The ability to compute theoretical mode shapes for the vertical and radial components of Rayleigh waves is particularly beneficial for evaluating modal amplitudes and identifying the predominant mode. The theoretical framework of the TLM has been extensively documented in the literature (Kausel and Roësset 1981; Kausel 1999; Barbosa and Kausel 2012; Bhaumik and Naskar 2024a). Therefore, only the key steps relevant to the present formulation are summarized here. In particular, emphasis is given to the theoretical computation of vertical and radial modal responses at the surface, which are intrinsic modal properties of the layered medium and independent of source-receiver orientation or acquisition geometry. The source-receiver directivity effects are subsequently incorporated through the combination of the corresponding component-specific amplitudes required for defining the radial-component predominant-mode.

Rayleigh-wave propagation in a layered half-space can be formulated as a quadratic eigenvalue problem using the TLM in the frequency-wavenumber domain (Kausel and Roësset 1981):

$$[k^2\mathbf{A} + ik\mathbf{B} + (\mathbf{C} - \omega^2\mathbf{M})]\{\boldsymbol{\phi}_R\} = 0, \tag{1}$$

where the $\omega$ is angular frequency, $k$ is the wavenumbers, $\boldsymbol{\phi}_R$ is the right eigenvector representing mode shapes, and $i = \sqrt{-1}$. The global stiffness matrices $\mathbf{A}$, $\mathbf{B}$, $\mathbf{C}$ and $\mathbf{M}$ are assembled from contributions of individual thin layer elements that discretize the subsurface layers in the vertical direction. The resulting displacement field consists of coupled vertical and radial components, reflecting the elliptical particle motion characteristic of Rayleigh waves. For computational efficiency, the quadratic equation is reformed into an equivalent generalized eigenvalue problem, allowing the use of standard solution procedures. Accordingly, the global system is subdivided into horizontal ($x$) and vertical ($z$) components. By arranging the horizontal components degrees of freedom first, followed by vertical components, the quadratic eigenvalue problem can be transformed into an equivalent generalized eigenvalue form (Kausel and Peek 1982):

$$\left(k^2\begin{bmatrix}\mathbf{A}_{xx} & 0\\ \mathbf{B}_{xz}^T & \mathbf{A}_{zz}\end{bmatrix}+\begin{bmatrix}\mathbf{C}_{xx}-\omega^2\mathbf{M}_{xx} & \mathbf{B}_{xz}\\ 0 & \mathbf{C}_{zz}-\omega^2\mathbf{M}_{zz}\end{bmatrix}\right)\begin{Bmatrix}\boldsymbol{\phi}_{\mathbf{x}}\\ ik\boldsymbol{\phi}_{\mathbf{z}}\end{Bmatrix}=0 \tag{2}$$

where $\phi_x and\, \phi_z$ represent horizontal and vertical displacement components of the eigenvector, respectively. $\mathbf{A}_{xx}$, $\mathbf{A}_{zz}$, $\mathbf{B}_{xz}$, $\mathbf{C}_{xx}$, $\mathbf{C}_{zz}$, $\mathbf{M}_{xx}$ and $\mathbf{M}_{zz}$correspond to the subdivided contributions of the global matrices (Kausel and Peek 1982). Solving this system yields the wavenumbers and their associated mode shapes. The eigenvectors obtained from the solution are then normalized to satisfy the modal orthogonality condition (Barbosa and Kausel 2012). The normalized eigenvectors are expressed as, $\widehat{\boldsymbol{\Phi}} = \mathbf{N}_R\boldsymbol{\Sigma}^{-1/2}$, where $\boldsymbol{\Sigma} = \mathbf{N}_L^T\overline{\mathbf{A}}\,\mathbf{N}_R\boldsymbol{\lambda}^{-1/2}$, $\mathbf{N}_R$ and $\mathbf{N}_L$ are associated with right and left eigenvector matrices, $\overline{\mathbf{A}} = [\mathbf{A}_{xx} \quad \mathbf{0}\,; \mathbf{B}_{xz}^T \quad \mathbf{A}_{zz}]$, and $\boldsymbol{\lambda} = \mathrm{diag}\{k_1^2, k_2^2, \ldots\}$ is the matrix containing eigenvalues (Barbosa and Kausel 2012; Astaneh and Guddati 2016). The normalized eigenvectors are subsequently used to evaluate surface amplitudes of the vertical and radial components, and then those amplitudes are combined to calculate the predominant mode of the radial component due to a vertical source.

At the free surface ($z = 0$), the normalized vertical component of the $m^{\text{th}}$ mode at frequency $\omega$, is written as: $\widehat{\boldsymbol{\Phi}}_{z,m}(0;\ k_m, \omega)$, representing the theoretical surface amplitude associated with that mode. Similarly, the normalized horizontal component is represented by $\widehat{\boldsymbol{\Phi}}_{\boldsymbol{x},m}(0;\ k_m, \omega)$. These modal amplitudes are intrinsic properties of the layered medium. However, the contribution of each surface-wave mode to the recorded wavefield is governed by a modal participation that accounts for both source and receiver directivity (Kausel and Peek 1982). The excitation strength of a given mode depends on the product between its eigenfunction and the applied source polarization. Accordingly, predominant modes for different source-receiver configurations can be defined from the product of the corresponding modal components.

Assuming vertically polarized source excitation and vertical-component measurements, the predominant mode for the vertical-source with vertical-receiver (V–V) configuration is given by:

$$k_{predominant}^{(V-V)}(\omega) = \arg\max_{k_m}\left|\widehat{\boldsymbol{\Phi}}_{z,m}(0;\ k_m, \omega)\,\widehat{\boldsymbol{\Phi}}_{z,m}(0;\ k_m, \omega)\right| \tag{3}$$

Note that, for the V–V configuration, the predominant mode obtained from $\left|\widehat{\boldsymbol{\Phi}}_{z,m}\widehat{\boldsymbol{\Phi}}_{\boldsymbol{z},m}\right|$ is identical to that obtained using only the vertical modal amplitude $\left|\widehat{\boldsymbol{\Phi}}_{z,m}\right|$ (Bhaumik and Naskar 2024b; Bhaumik and Cox 2026), since squaring preserves the relative modal dominance.

Similarly, the predominant mode for the radial-source with radial-receiver (R–R) configuration can be expressed as:

$$k_{predominant}^{(R-R)}(\omega) = \arg\max_{k_m}\left|\widehat{\boldsymbol{\Phi}}_{\boldsymbol{x},m}(0;\ k_m, \omega)\,\widehat{\boldsymbol{\Phi}}_{\boldsymbol{x},m}(0;\ k_m, \omega)\right| \tag{4}$$

Assuming the wavefield is dominated by vertical sources and radial-component measurements, the predominant mode for the vertical-source with radial-receiver (V–R) configuration is obtained from the product of the vertical and horizontal modal amplitudes:

$$k_{predominant}^{(V-R)}(\omega) = \arg\max_{k_m}\left|\widehat{\boldsymbol{\Phi}}_{z,m}(0;\ k_m, \omega)\,\widehat{\boldsymbol{\Phi}}_{\boldsymbol{x},m}(0;\ k_m, \omega)\right| \tag{5}$$

The corresponding quantity $\left|\widehat{\boldsymbol{\Phi}}_{z,m}\widehat{\boldsymbol{\Phi}}_{\boldsymbol{x},m}\right|$ represents a simplified modal participation amplitude associated with the V–R configuration and forms the basis of the proposed RCPM formulation. Note that source-receiver directivity plays a significant role in controlling the modal energy distribution observed in

dispersion images. For instance, an in-line horizontal source would generate a different modal response, leading to a different predominant mode compared to that obtained from a vertical source. However, in practical field applications, vertical sources are more commonly used due to ease of implementation and consistency in excitation. Therefore, the present study focuses on the V–R predominant mode for the RCPM formulation.

It should be emphasized that the proposed predominant-mode formulation is conceptually different from an effective or superimposed mode formulation (Lai 1998; Foti 2000; Lai et al. 2014; Bhaumik and Naskar 2026). Instead of representing the total wavefield through summation of all modal contributions, the present approach identifies the mode with the highest modal participation at each frequency. This formulation provides a practical advantage for surface-wave inversion, as it does not require detailed information regarding source location, receiver geometry, or array length, which are required for an effective mode formulation. In practical applications, dispersion targets are often constructed by combining dispersion data from multiple active shots acquired from both sides of the array (Vantassel and Cox 2022; Yust et al. 2024), and in many cases including passive measurements to increase investigation depth. Under such conditions, obtaining a single effective mode for acquisition configurations with different geometries is challenging. The proposed predominant-mode formulation provides a simplified, yet physically interpretable, representation of the dominant modal behavior without requiring explicit modeling of these acquisition-dependent effects.

## 3. Dispersion Behavior of Vertical and Radial Components of Rayleigh Waves

In this section, the dispersion behavior of Rayleigh waves obtained from the vertical and radial components is examined. To investigate the modal energy distribution associated with these two components, three representative synthetic ground models are considered (Figure 1): (i) Ground Model I: a $V_s$ model consistently increasing with depth, (ii) Ground Model II: a $V_s$ model containing a low-velocity layer (LVL), and (iii) Ground Model III: a $V_s$ model containing a high-velocity layer (HVL). Ground Model I (Figure 1a) is adopted from Vantassel and Cox (2022) and it is characterized by an 'L'-shaped dispersion curve, which has been shown to be difficult to invert accurately. The same model was further analyzed by Bhaumik and Cox (2026) using a vertical-component predominant-mode inversion, where modal osculation at low frequencies was observed due to a strong velocity contrast near 30 m. The LVL and HVL profiles (Figure 1b and 1c) are derived from Ground Model I by modifying the $V_s$ and thickness of the upper three layers. These models are selected to represent typical subsurface conditions that are known to influence modal behavior and energy distribution in Rayleigh-wave dispersion.

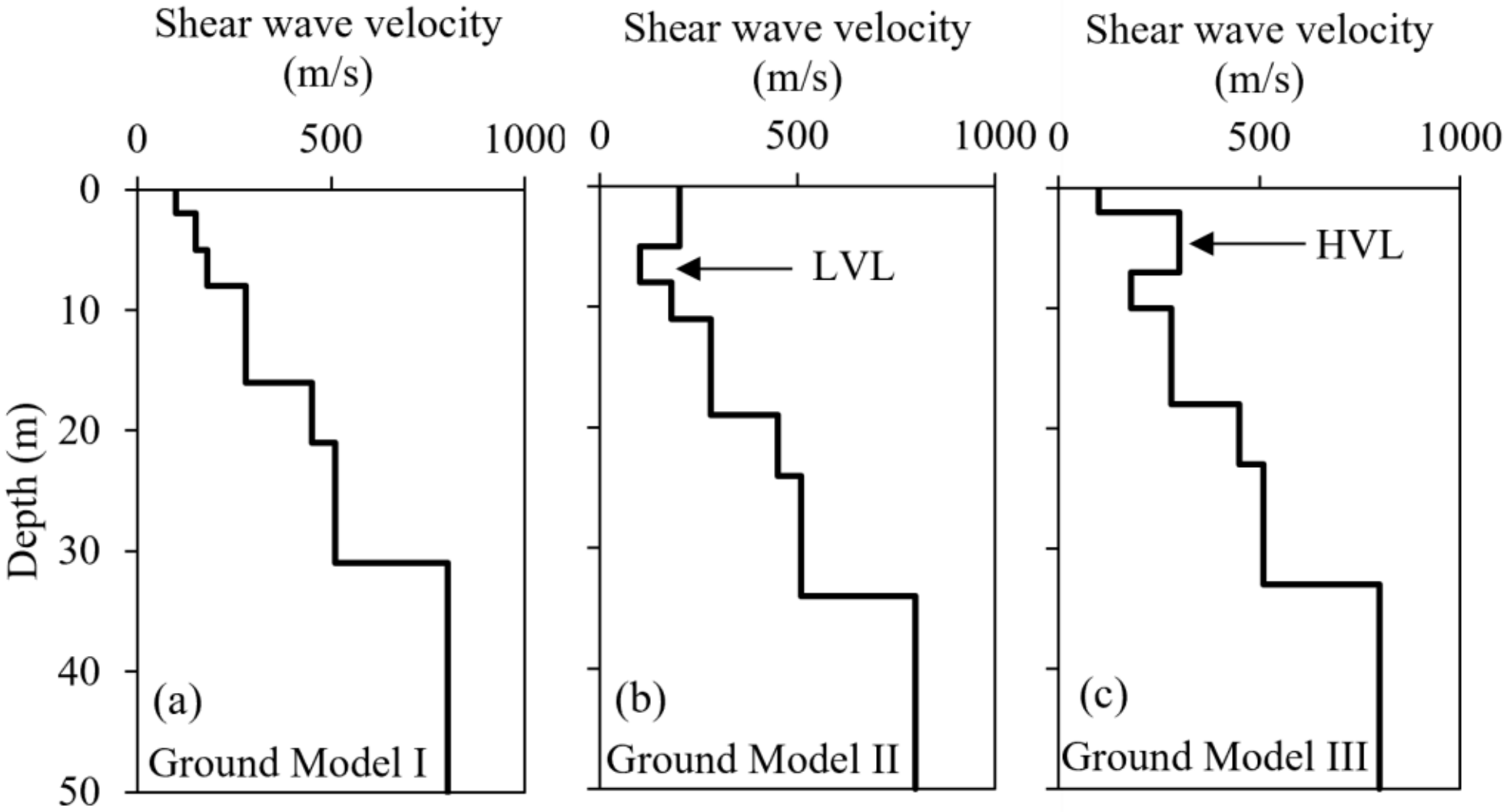

Figure 1. $V_S$ models used in this study: (a) Ground Model I: monotonically increasing velocity model, (b) Ground Model II: with a low-velocity layer (LVL), and (c) Ground Model III: with a high-velocity layer (HVL).

Figure 2 presents the theoretical dispersion curves for the three synthetic models overlaid with normalized modal amplitudes at the surface, enabling a direct comparison of modal energy distribution across components. For illustration purpose, the amplitudes are normalized at each frequency to make the maximum modal amplitude equal to 1.0. It is emphasized that this frequency-wise normalization is applied solely for visualization and is distinct from the matrix normalization procedure described in the previous section. Figures 2(a–c) illustrate the dispersion curves superimposed with normalized theoretical modal participation amplitudes at the surface for the vertical-source vertical-receiver (V–V) configuration. Figures 2(d-f) show the corresponding normalized modal participation amplitudes for the radial-source radial-receiver (R–R) configuration. Figures 2(g–i) present the normalized modal participation amplitudes for the vertical-source radial-receiver (V–R) configuration. The comparison highlights that modal energy distribution is strongly dependent on the measurement component, the source directionality, and the underlaying subsurface stratigraphy, primarily due to the elliptical particle motion characteristic of Rayleigh waves.

For Ground Model I, the V–V configuration (Figure 2a) exhibits a shift in modal energy from R0 to R1 near the osculation point (OP) at 7 Hz, whereas the R–R configuration (Figure 2d) shows dominance of R0 across the entire frequency range. This type of reduction in the R0 vertical amplitude often leads to peaks in the ellipticity curve. However, when the radial response due to a vertical source (V–R configuration) is considered (Figure 2g), a transition in modal energy from R0 to R1 is observed at a lower frequency than the OP of the V–V configuration. Therefore, although modal osculation is an intrinsic property of the theoretical dispersion curves (Kausel et al. 2015), its appearance in the dispersion images depends on modal amplitude and measurement configuration. Overall, the V–R predominant mode shows behavior closer to the V–V configuration than the R–R configuration across all frequencies. This highlights the importance of incorporating source–receiver configuration when interpreting radial component dispersion characteristics.

For Ground Model II (Figure 1b), the presence of a LVL significantly alters the modal energy distribution relative to Ground Model I. In the V–V configuration (Figure 2b), strong modal interaction between R0 and R1 is still observed at low frequencies, with clear energy redistribution between modes near the OP at 5 Hz. At higher frequencies, the influence of the LVL leads to additional redistribution of energy, where modes R2 and R3 are excited while mode R1 is comparatively suppressed, resulting in a complex dispersion pattern. The R–R configuration (Figure 2e) exhibits even more erratic mode behavior at high frequencies but fails to completely transition to R1 below the OP at low frequencies. Overall, for Ground Model II the V–R configuration (Figure 2h) shows behavior closer to the V-V configuration at higher frequencies, but closer to the R-R configuration at lower frequencies.

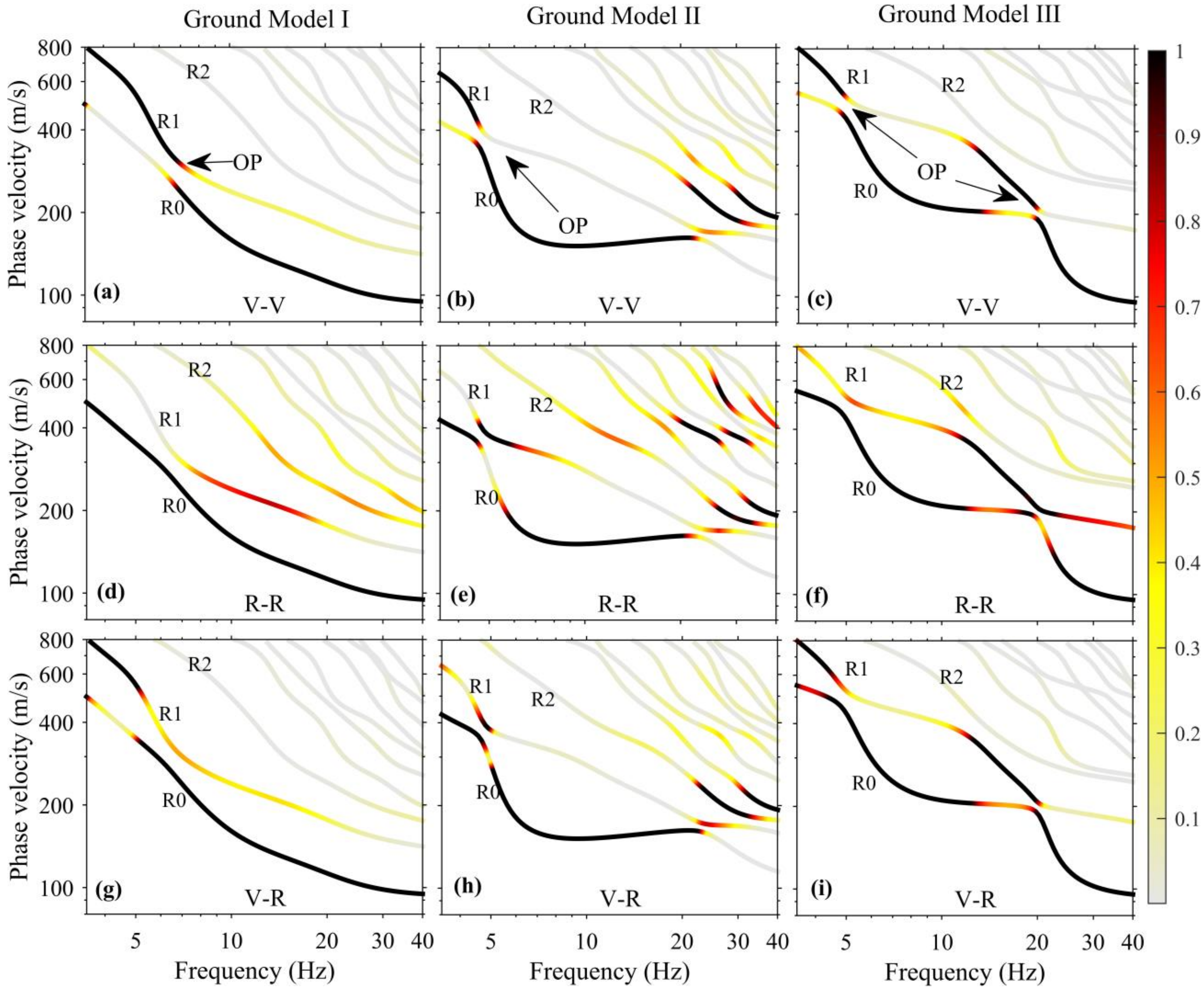

Figure 2. Dispersion curves overlaid with normalized modal amplitudes at the surface for three representative profiles: Ground Model I with increasing velocity (first column), Ground Model II with low-velocity layer (LVL, second column), and Ground Model III with high-velocity layer (HVL, third column). The top row (a–c) shows the vertical-source with vertical-component (V-V) amplitudes, the middle row (d–f) shows radial-source with radial-component (R-R) amplitudes, and the bottom row (g–i) shows vertical-source with radial-component (V-R) amplitudes. The color scale represents normalized modal energy.

For Ground Model III (Figure 1c), the modal behavior is influenced by the presence of a HVL near the surface combined with the strong velocity contrast at depth. In the V–V configuration (Figure 2c), the R0 and R1 modes exhibit multiple regions of energy redistribution, producing a pattern analogous to a double osculation scenario. Specifically, modal energy transfers from R0 to R1 at two distinct OPs: first, approximately 20 Hz, and again near approximately 5 Hz. In contrast, the R–R configuration (Figure 2f) shows a noticeably different modal energy distribution at higher frequencies, and a lack of complete modal energy transition to R1 at lower frequencies. For this case, the V–R configuration (Figure 2i) exhibits an energy distribution more comparable to the V–V, once again highlighting the importance of considering the source mechanism when evaluating the radial component predominant mode.

The three ground models discussed above highlight the complexity in modal energy behavior arising from different subsurface stratigraphy and demonstrate that the modal energy for a radial component due to a vertical load can differ significantly from those associated with the V–V and R–R configurations. Such variations have important implications for surface-wave inversion, as the identified dispersion data may not correspond to the same modal order across different measurement components. Similar discrepancies in vertical and radial dispersion images, obtained using different geophone components from a common vertical source, were reported by Vantassel and Cox (2021), which can be attributed to the component-dependent modal behavior observed in this study. These findings further emphasize the need to explicitly account for measurement direction when interpreting surface wave dispersion data and developing inversion frameworks.

To address the use of radial component wavefields to obtain dispersion data and the ambiguities that can arise when attempting manual modal indexing of the dispersion data during inversion, this study proposes a RCPM (radial-component predominant-mode) inversion approach derived from Equation (5). The proposed RCPM framework addresses two key challenges: (i) it eliminates the need for explicit modal identification by inherently incorporating modal energy dominance during the dispersion-misfit calculation, and (ii) it is specifically formulated for the radial component of Rayleigh waves emanating from a vertical source, making it directly applicable to DAS-based measurements. This formulation forms the basis of the inversion framework presented in the following section.

# 4. Inversion Framework: Synthetic Examples

This section presents the implementation of the proposed RCPM inversion framework to the three synthetic ground models discussed above. The formulation of inversion targets and selection of inversion model parameters are first described, followed by the implementation of a differential evolution (DE)–based global optimization scheme. We note that the implementation in a DE-based global optimization is simply to demonstrate the applicability of the approach, and that the RCPM forward problem could be implemented in other inversion optimization schemes as desired.

## 4.1. Development of Inversion Targets

The inversion target is constructed by mimicking typical surface-wave acquisition practices associated with incorporating both active and passive measurements to generate dispersion data over a broader frequency bandwidth. Active-source data are simulated using three shot locations positioned at offsets of 5, 10, and 15 m from the receiver array, as would be typical for multi-channel analysis of surface waves (MASW) testing. To represent passive data acquisition, additional sources were placed at large distances (approximately 1–2 km from the array), thereby approximating a far-field condition. The simulated time-domain data were generated using the wavefield modeling approach described in Bhaumik and Naskar (2024), employing a vertical source and recording the radial component of Rayleigh waves. For active-source data sets, 24 channels with a separation of 2 m were used, whereas 100 channels at 2 m separation were considered for the passive configuration to ensure sufficient spatial coverage and resolution of low-frequency content. These channel numbers and channel separations are easy to obtain with many current DAS interrogator units. The passive sources at large distances were positioned in-line with the linear measurement array. Therefore, for both active and passive configurations, wavefield dispersion processing was performed using the *SWprocess* workflow (Vantassel and Cox 2022) and the frequency domain beamforming (FDBF) approach (Zywicki and Rix 2005). In real-field passive data recorded with a linear array, cross-correlation of noise waveforms is often performed prior to dispersion processing by treating one channel as a virtual source; however, in the present synthetic framework this was not required, as sources were explicitly defined in the waveform simulations. Nonetheless, it is important to note that the use of a linear array for real-field passive (ambient noise) measurements may introduce uncertainty in the dispersion-data due to the unknown direction of ambient noise wave propagation. In practice, ambient noise wavefields may not be propagating perfectly in-line with the array, and waves arriving from off-line directions can introduce high apparent phase velocities (Cox and Beekman 2011). Although DAS measurements are most sensitive to motion aligned with the fiber direction, thereby reducing sensitivity to off-line arrivals, such effects cannot be fully eliminated. Therefore, some degree of uncertainty is expected when interpreting real-field dispersion data derived from linear-array passive measurements and readers should be aware of this.

Dispersion data were extracted from the active and passive wavefield simulations by selecting the dominant energy trends visible after applying the FDBF wavefield transformation. The active and passive dispersion data were then combined to form a statistical dispersion target for each ground model, as shown in Figure 3. As dispersion data uncertainty from simulated waveforms tends to be under-predicted relative to actual field waveforms, a minimum coefficient of variation (COV) of 0.05 was adopted. Note that the dispersion data extracted from the simulated waveforms are consistent with the predominant energy trends of a radial-component and a vertical source, as shown in Figure 2(g-i), and also plotted in the background of Figure 3. This agreement supports both the validity of the proposed analytical formulation as well as the waveform simulations, particularly in capturing modal dominance and transitions.

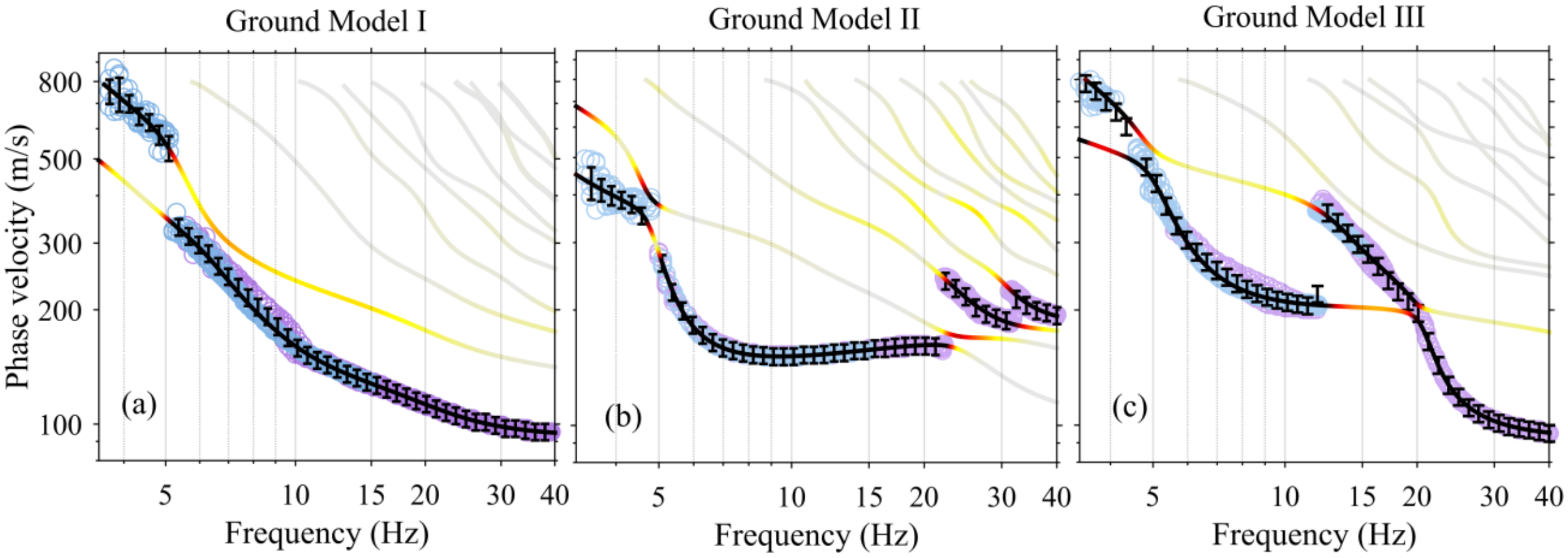

Figure 3. Statistical dispersion targets for Ground Models I–III, shown in panels (a), (b), and (c), respectively. Dispersion data derived from simulated active sources (purple markers) and passive sources (blue markers) are merged. Error bars represent dispersion data uncertainty from the simulated waveforms and a minimum COV of 0.05. The theoretical modal dispersion curves for a vertical source and radial component are plotted in the background.

## 4.2. Inversion Procedure

The inversion was carried out using the Differential Evolution algorithm (Storn and Price 1997), a population-based global optimization technique suitable for highly nonlinear and multimodal problems such as surface-wave inversion. DE iteratively evolves a population of candidate models through mutation, crossover, and selection operations, enabling efficient exploration of the model space without requiring gradient information. At each iteration, trial models are generated by combining randomly selected individuals from the current population and are accepted if they yield an improved objective function value. An adaptive control parameter and a $p$-best mutation strategy is used to improve convergence. An initial population of $N_p$ candidate models is randomly generated within the model parameter bounds. At each generation, models are ranked based on their objective function value, and a subset corresponding to the top $p$ =15% of the population is identified as the $p$-best. Mutation is performed using a "current-to-$p$best/1" scheme (Zhang and Sanderson 2009), where each parameter vector $\mathbf{x}_i^{(g)}$ at generation $g$ is perturbed according to:

$$\boldsymbol{v}_i^{(g)} = \mathbf{x}_i^{(g)} + F_i\left(\mathbf{x}_{pbest}^{(g)} - \mathbf{x}_i^{(g)}\right) + F_i\left(\mathbf{x}_{r1}^{(g)} - \mathbf{x}_{r2}^{(g)}\right), \tag{6}$$

where $\boldsymbol{v}_i^{(g)}$ is the muted parameter vector, $\mathbf{x}_{pbest}^{(g)}$is randomly selected from the $p$-best models, and $r1$ and $r2$ are indices. In the present implementation, the mutation factor $F_i$ was adaptively varied across generations to balance exploration and exploitation, decreasing linearly from $F_{\max} = 0.8$ to $F_{\min} = 0.3$. The objective misfit function was defined as a weighted root-mean-square dispersion misfit ($m_d$) based on the difference between observed and calculated dispersion curve, incorporating uncertainty in the measurements (Wathelet 2005):

$$m_d = \sqrt{\sum_{i=1}^{N_f} \frac{\left(\boldsymbol{V}_{\mathrm{obs},i} - \boldsymbol{V}_{\mathrm{predominant},i}\right)^2}{N_f \sigma_i^2}} \tag{7}$$

where $\boldsymbol{V}_{\mathrm{obs},i}$ is the observed phase velocity at frequency $i$, $\boldsymbol{V}_{\mathrm{predominant},i}$ is the modeled radial-component predominant-mode velocity at the same frequency obtained from Equation (4), $N_f$ is the total number of frequency points, and $\sigma_i$ is the standard deviation associated with the observed data at frequency $i$. Note that the proposed inversion framework is flexible, and the forward modeling scheme is independent of the specific optimization algorithm employed. Although the present study adopts DE due to its robustness in handling nonlinear and multimodal objective functions, the framework can readily accommodate other global optimization strategies (e.g., Genetic Algorithms, Particle Swarm Optimization, Bayesian approaches, or gradient-based methods where applicable) without modification to the forward solver.

## 4.3. Inversion Model Parameterization

An appropriate parameterization of trial subsurface models is necessary to balance resolution and stability during inversion. While the true number of layers is known for each synthetic ground model, we wanted to treat the inversion layering parameterizations as if we did not know the true number of layers, thereby reflecting more challenging and realistic conditions. In the absence of a priori information on the true number of subsurface layers, a systematic approach based on layering ratios (LR), as proposed by Cox and Teague (2016) and implemented in the *SWinvert* workflow (Vantassel and Cox 2021), can be used to investigate several layering parameterizations. In the LR approach, the thickness of the first layer is constrained based on the minimum observed wavelength to ensure adequate resolution of shallow layers, while the thickness of subsequent layers is encouraged, but not forced, to increase with depth according to a selected LR (Cox and Teague 2016). A range of LR values (1.25, 1.5, and 2.0) was considered in the synthetic study to generate multiple candidate subsurface layering discretizations, thereby avoiding dependence on a single layering scheme and allowing exploration of model uncertainty. The $V_s$ bounds for each layer were defined based on the observed dispersion data. The lower bound was set at half of the minimum phase velocity, while the upper bound was selected as 25% greater than the maximum phase velocity (1000 m/s in this case), following Vantassel and Cox (2021). Mass density was assumed to be constant at 2000 kg/m$^3$, and Poisson's ratio was allowed to vary within the range of 0.25-0.35, representing un-saturated subsurface conditions. This parameterization is consistent with the depth-dependent resolving capability of surface waves and provides a controlled, yet flexible, framework for inversion. Note that the adopted inversion parameterization represents one reasonable practical choice for the present study; however, alternative parameterizations may be adopted depending on site conditions and prior information.

# 5. Inversion Results for Synthetic Examples

This section evaluates the performance of the proposed RCPM inversion framework using the three synthetic ground models. To demonstrate its effectiveness, the proposed approach is compared with CMM inversion strategies. The CMM inversions were performed using the *Dinver* module in *Geopsy* (Wathelet 2008). In the CMM approach, modal indexing is explicitly assigned where obvious features such as clear separation or mode jumps are observed in the target dispersion data. In cases where such scenarios are ambiguous or absent, multiple possible modal interpretations are considered, and the *Dinver* toolbox is allowed to associate modes based on minimum misfit. As the conventional inversion relies on subjective interpretation of an analyst, multiple scenarios are evaluated to reflect the range of possible analyst-driven decisions.

Following the procedure outlined in the *SWinvert* workflow (Vantassel and Cox 2021), each selected LR parameterization is evaluated through 10 independent inversion trial runs as a means to account for uncertainty. For each LR, the 10 best-fitting subsurface models from each run, along with the overall best-fit model, are retained for comparison. In addition to visual comparison of the inverted $V_s$ profiles, a quantitative statistical assessment is performed using the dispersion misfit (Equation 7), representing the goodness of data fit, and the $V_s$ misfit ($m_{V_s}$) (Equation 8), which evaluates inversion accuracy relative to the known $V_s$ profiles of the synthetic models. These metrices are summarized using bar plots, with the lower and upper bounds representing the minimum and maximum values from the 10 independent trials. Furthermore, the profile corresponding to the minimum dispersion misfit is examined in terms of its associated $V_s$ misfit to assess the consistency between data fitting and model accuracy.

The $m_{V_s}$ values between the true synthetic $V_s$ profiles and the inverted $V_s$ profiles are computed up to a depth of 80 m, with a depth discretization of 0.5 m, using mean absolute relative error (Vantassel and Cox 2021):

$$m_{V_s} = \frac{1}{N}\sum_{j=1}^{N_d} \frac{\left|\boldsymbol{V}_{s_j},\text{inverted} - \boldsymbol{V}_{s_j},\text{true}\right|}{V_{s_j},\text{true}} \tag{8}$$

where $V_{s_j},\text{inverted}$ and $V_{s_j},\text{true}$ are the inverted and true shear wave velocities at depth point $j$, respectively, and $N_d$ is the number of depth discretization points. The evaluation depth of 80 m is selected based on the resolution limit associated with the maximum wavelength in the observed/simulated dispersion data (approximately one-half to one-third of the wavelength) and is applied consistently across all models to enable direct comparison.

### 5.1. Ground Model I: Increasing velocity

Ground Model I is a 6-layer increasing velocity profile, with the deepest $V_s$ boundary at 31 m overlaying an elastic half-space (Figure 1a). Vantassel and Cox (2021) classified this model as challenging to invert accurately. Bhaumik and Cox (2026) further showed that the model exhibits higher mode excitation at low frequency due to modal osculation and demonstrated that it can be reasonably retrieved through vertical-component predominant-mode inversion of vertical component dispersion data. However, as shown in Figures 2a, 2d, and 2g, the dispersion data energy/amplitude trends differ considerably when considering vertical and radial components, motivating the need for a component-consistent inversion framework if extracting dispersion data from radial components due to a vertical source.

Figure 4 presents the inversion results for the radial component dispersion target data shown in Figure 3a using the proposed RCPM and a CMM inversion. The corresponding LR = 1.25, 1.5, and 2.0 parameterizations consist of 12, 8, and 5 layers, respectively. For the CMM inversion, the data above 5 Hz was interpreted as R0 (see Figure 3a) and below 5 Hz as R1 due to the clearly visible discontinuity near this frequency. Figure 4a compares the simulated target dispersion data with theoretical modal curves obtained from RCPM and CMM inversions. The proposed RCPM fits the data well and implicitly selects the appropriate modal branch based on energy dominance, without requiring explicit mode indexing. As shown in Figure 5a, the RCPM yields lower dispersion misfit, $m_d$ compared to CMM. The CMM inversion achieves comparable, but slightly higher, fits to the data; however, it requires explicit modal interpretation by the analyst. Figure 4b shows that both the RCPM and CMM inversions capture the overall velocity trend, with some scatter in estimating the depth of the deepest interface, which is consistent with the reduced resolution of surface waves at greater depths (Vantassel and Cox 2021). Notably, there is greater scatter in the inverted $V_s$ profiles for the CMM approach. This can also be observed by comparing the $V_s$ misfit, $m_{V_s}$ values shown in Figure 5b, wherein the ranges for the $V_s$ misfit values are higher for the CMM inversions than for the RCPM inversions. Figure 4c compares the single best-fit $V_s$ profiles for each LR parameterization. Note that the lower bounds of the $V_s$ misfit bars in Figure 5b are similar for both the methods, indicating that each method has solutions close to the true model within the ensemble of 100 profiles (10 per trial) shown in Figure 4b. However, the $V_s$ misfits corresponding to the single best-fit (i.e., lowest $m_d$) profiles plotted in Figure 4c, which are typically selected in practice, do not consistently stay near the lower bound of the $m_{V_s}$ distributions (indicated by the black stars in Figure 5b). This suggests that selecting models solely based on dispersion misfit does not guarantee optimal recovery of the true velocity structure. In this regard, the proposed RCPM inversion demonstrates improved performance relative to the CMM inversions, as it has the lesser amount of scatter in the $V_s$ misfit values and its lowest dispersion misfit models have lower $V_s$ misfit values compared to CMM for all LR parameterizations. Overall, this detailed analysis highlights the importance of employing a component-consistent forward model, as the proposed RCPM framework not only achieves the best fits to the experimental data, but also improves the reliability and consistency of the inverted subsurface $V_s$ profiles.

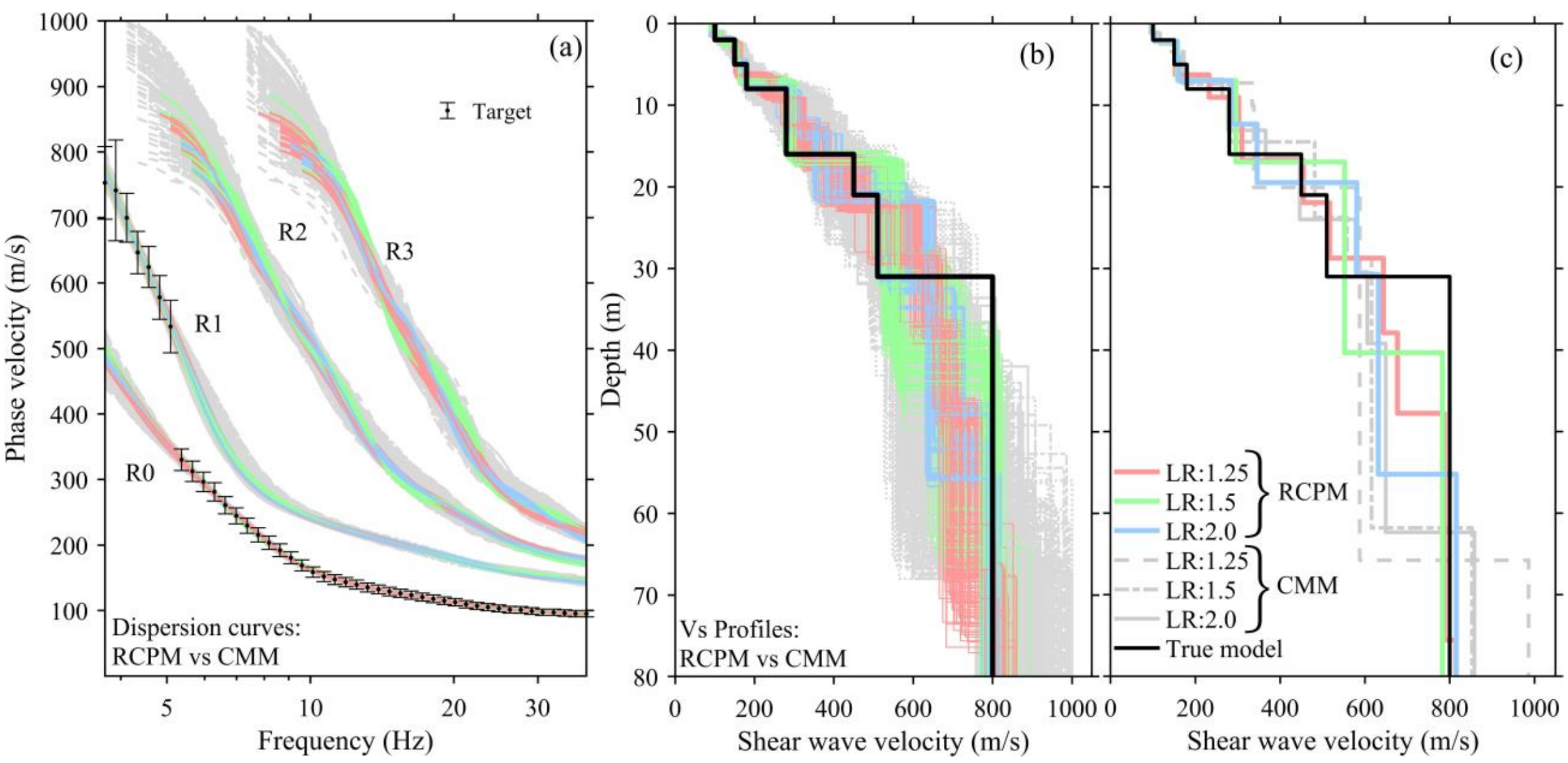

Figure 4. Inversion results for Ground Model I using RCPM and CMM: (a) target dispersion data (black points with uncertainty bars) and derived theoretical dispersion curves (R0-R3) corresponding to the 100 best-fit inverted profiles per LR (ten per trial), (b) ensemble of inverted $V_s$ profiles for different layering ratios (LR = 1.25, 1.5, and 2.0), along with the true model, and (c) comparison of inverted best-fit $V_s$ profiles per LR.

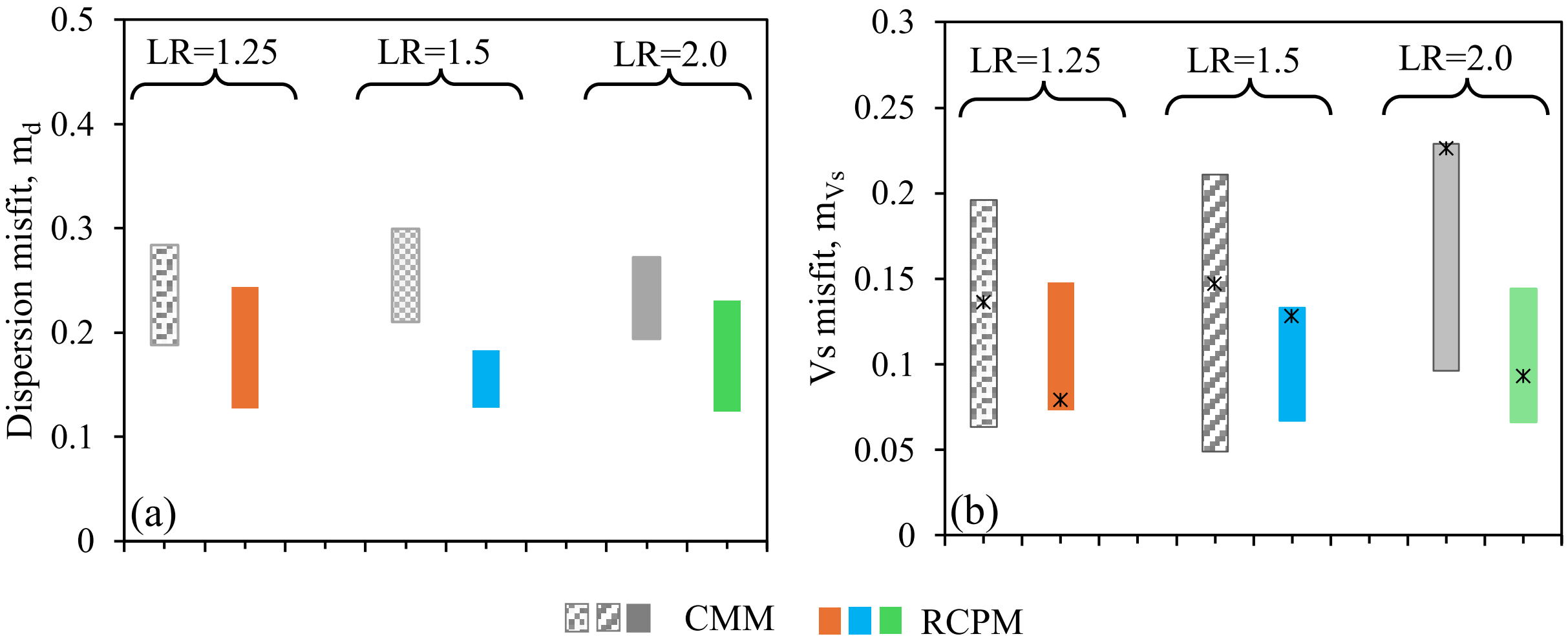

Figure 5. Comparison of RCPM and CMM inversion performance for Ground Model I: (a) ranges in dispersion misfit ($m_d$), and (b) ranges in shear wave velocity misfit ($m_{V_s}$) obtained from the 100 best-fit inverted profiles per LR (ten per trial). The bar ranges represent the minimum and maximum values across trials, with black asterisk markers in the $m_{V_s}$ bars indicating the lowest dispersion misfit profiles.

### 5.2. Ground Model II: LVL

The presence of LVL in Ground Model II (Figure 1b), introduces significant ambiguity in modal identification and indexing, as discussed in Figure 2b and 2h, making the conventional inversion challenging. Figure 6 presents the inversion results for the radial component dispersion target data shown in Figure 3b using the proposed RCPM and CMM approaches. For the CMM inversion, two practical scenarios were considered based on the dispersion trends in Figure 3b. In Scenario I, the dispersion data below 20 Hz were treated as R0, with the two distinct higher frequency branches assigned to R1 and R2. This represents the most likely interpretation an analyst would make based on the apparent dispersion trends. In Scenario II, a more flexible interpretation was adopted: data below 5 Hz were assigned to either R0 or R1, the 5-20 Hz range was treated as R0, and higher the higher frequency branches were assigned to either R1 or R2 and either R2 or R3, respectively. This flexibility accounts for the possibility that higher modes may skip intermediate modes. When performing CMM inversions in *Geopsy* using target dispersion data that has been indexed with multiple possible modes (e.g., Scenario II), the most likely modal interpretations are determined based on minimizing the dispersion misfit values.

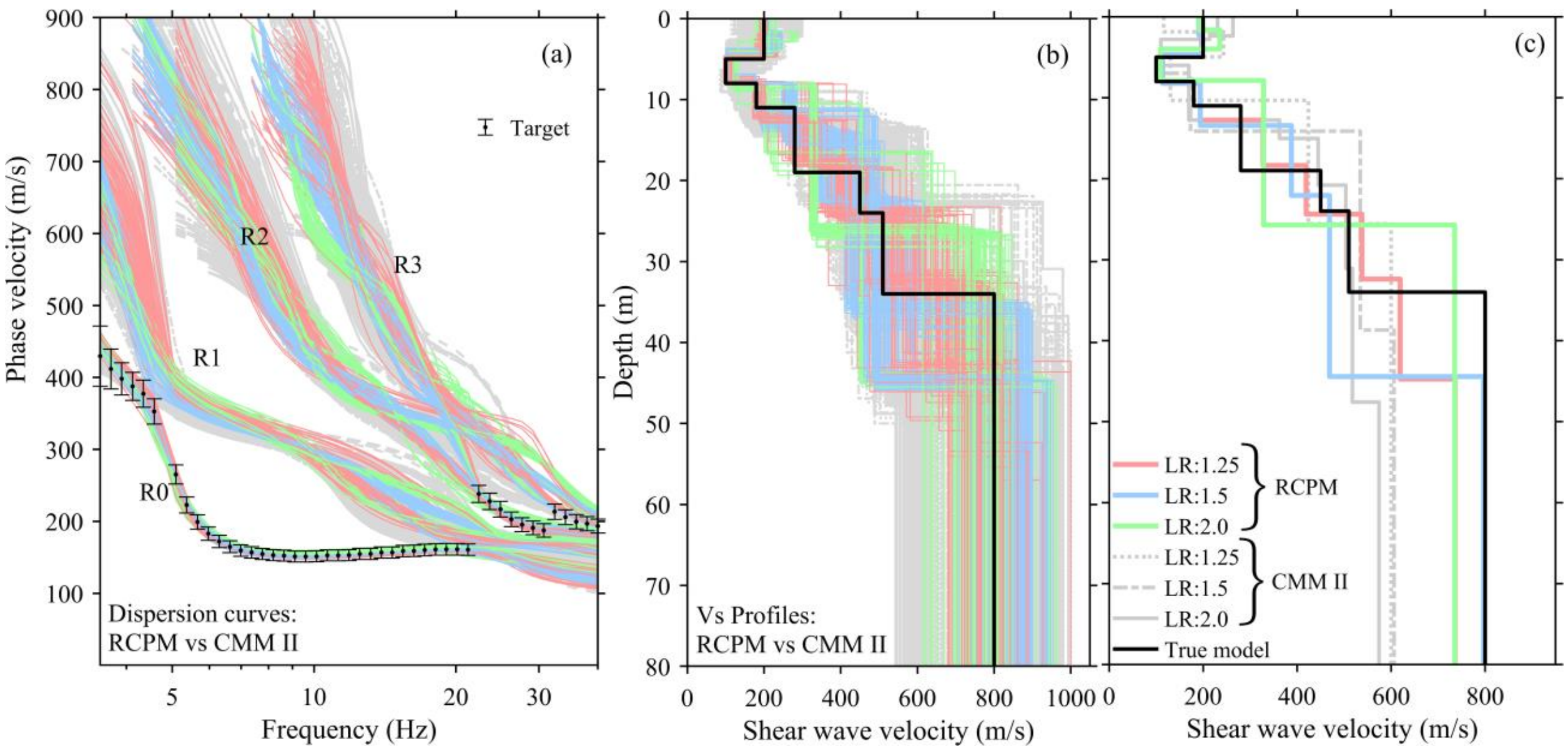

Figure 6. Inversion results for Ground Model II, using RCPM and CMM Scenario II: (a) target dispersion data (black points with uncertainty bars) and derived theoretical dispersion curves (R0-R3) corresponding to the 100 best-fit inverted profiles per LR (ten per trial), (b) ensemble of inverted $V_s$ profiles for different layering ratios (LR = 1.25, 1.5, and 2.0), along with the true model, and (c) comparison of inverted best-fit $V_s$ profiles per LR.

The inversion results for Ground Model II are provided in Figures 6 and 7. As the Scenario I CMM inversion produced comparatively higher dispersion misfits than Scenario II, its results are excluded from Figure 6 for clarity, although they are included in the statistical comparisons presented in Figure 7. Figure 6a compares the simulated radial component target dispersion data with theoretical dispersion curves obtained from the proposed RCPM and Scenario II CMM inversions. Both approaches provide good agreement with the target data and correctly skip the R1 mode. However, while the CMM achieves this through manual mode indexing by the analyst and frequency-wise misfit minimization, the RCPM identifies this by inherently accounting for modal energy distribution. Although the dispersion misfits of RCPM are higher than CMM (Figure 7a), the corresponding $V_s$ profiles shown in Figure 6b indicate that both methods capture the overall velocity trend. Notably, the CMM inversion exhibits greater scatter in the $V_s$ profiles, which can also be seen from the $V_s$ misfit distributions in Figure 7b. A comparison of the best-fit $V_s$ profiles across all LR values (Figure 6c) reveals that the CMM Scenario II inversions tend to underestimate the half-space velocity and misidentify layer interfaces below approximately 15 m depth. For CMM, although the lower bound of $V_s$ misfit is significantly better for Scenario II than Scenario I, the $V_s$ misfits associated with the best-fit profiles (indicated by black stars in Figure 7b) are similar in both scenarios, suggesting that either interpretation would yield a similar $V_s$ profile in practice if the model with the lowest dispersion misfit was chosen. In contrast, the proposed RCPM method yields consistently lower $V_s$ misfit values, both in terms of the lower bound and the best-fit profiles. The relatively higher dispersion misfits observed for LR = 2.0 can be attributed to the insufficient number of layers associated with this parameterization. In particular, the LR = 2.0 parameterization results in only four layers, compared to six layers in the true model and eight and six layers for LR = 1.25 and 1.5, respectively. Overall, the proposed RCPM inversion results indicate improved consistency and accuracy in recovering the subsurface $V_s$ profile despite having higher dispersion misfit values.

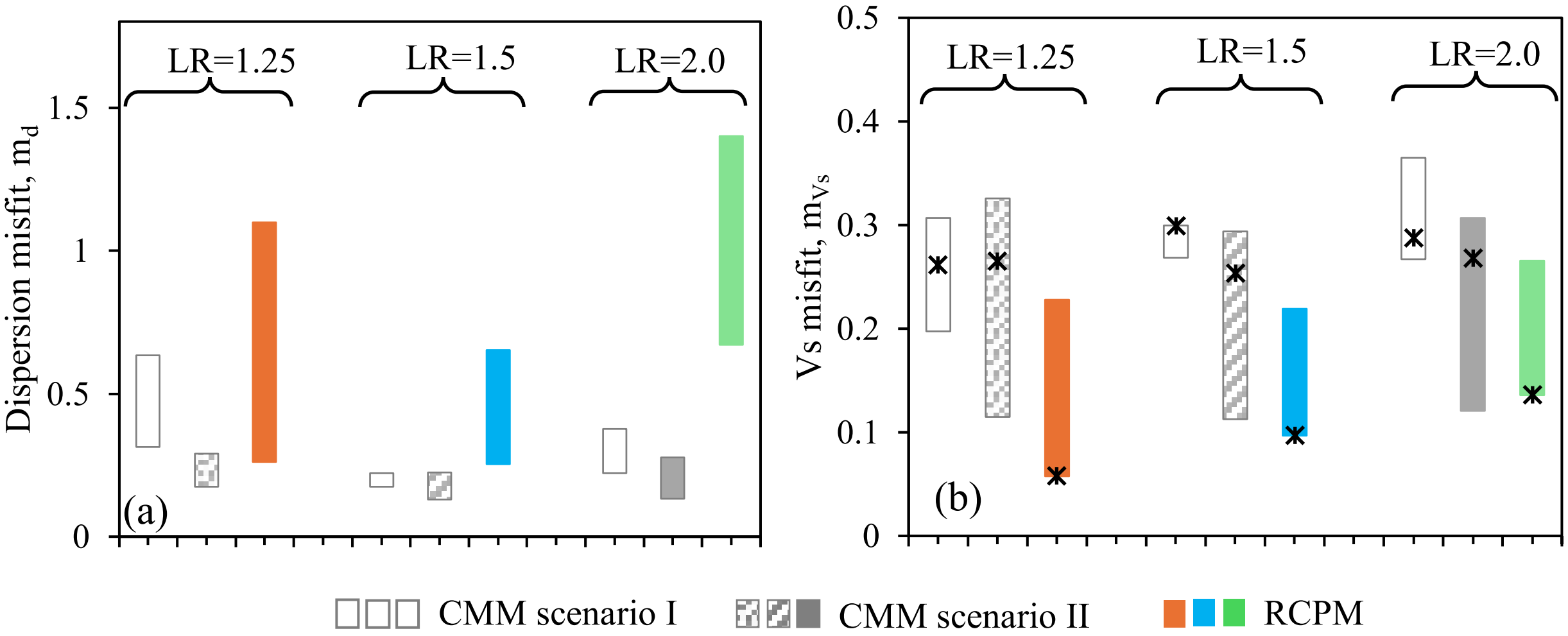

Figure 7. Comparison of RCPM and CMM inversion performance for Ground Model II: (a) ranges in dispersion misfit ($m_d$), and (b) ranges in shear wave velocity misfit ($m_{V_s}$) obtained from the 100 best-fit inverted profiles per LR (ten per trial). The bar ranges represent the minimum and maximum values across trials, with black asterisk markers in the $m_{V_s}$ bars indicating the lowest dispersion misfit profiles.

### 5.3.Ground Model III: HVL

In Ground Model III, as discussed earlier in Section 3 (Figures 2c, 2i and 3c), the presence of a HVL introduces a double-osculation-type behavior, complicating manual modal identification and indexing required for CMM inversions. Once again, two CMM inversion scenarios are considered to account for potential ambiguities in the model indexing. In Scenario I, all data below 12 Hz (Figure 3c) were treated as R0, as the discontinuity in the dispersion data near 5 Hz is not sufficiently clear to confidently assign it as R1. Data above 12 Hz were interpreted as either R0 or R1, as the transition frequency is not clearly observed. In Scenario II, a more flexible interpretation was adopted: the data below 5 Hz were assigned to either R0 or R1, the 5-12 Hz range was treated as R0, and the data above 12 Hz were assigned to either R0 or R1. The inversion results are presented in Figure 8, where the LR = 1.25, 1.5, and 2.0 parameterizations correspond to 13, 8, and 5 layers, respectively. For CMM, only Scenario II is retained for plotting in Figure 8 due to its comparatively better dispersion misfit than Scenario I (refer to Figure 9a). As shown in Figure 8a, the CMM Scenario II correctly identified the dispersion data below 5 Hz as the R1 mode, followed by the middle segment corresponding to R0. The higher-frequency segment (above 12 Hz) is variably interpreted, with portions assigned to either R0 or R1, and in some cases entirely assigned to R1. In contrast, the proposed RCPM approach consistently identifies the appropriate modal behavior without requiring explicit indexing. From Figure 9a, it can be observed that the dispersion-misfit values for all inversion approaches are broadly similar, with slightly lower values for the RCPM. Figure 8b shows the inverted $V_s$ profiles from the CMM Scenario II and RCPM inversions. Both approaches capture the overall trend of the subsurface profile, including the presence of a HVL; although, the inverted HVL layers are generally thinner than the HVL from the true profile, while the underlying LVL layers are generally thicker. These trends are more clearly observed in Figure 8c, which shows only the best-fit profiles across all LR values. While the dispersion misfit ranges for the two CMM inversion scenarios are nearly identical (Figure 9a), the corresponding $V_s$ misfit in Figure 9b is significantly higher for Scenario I, indicating that Scenario II provides more accurate subsurface estimates. In practical applications, where interpretation is often based primarily on dispersion misfit, such similarity in dispersion misfit for the two different CMM scenarios could lead to ambiguity and potential misinterpretation of subsurface $V_s$ profiles. Overall, the RCPM yields both the lowest dispersion misfit values and some of the lowest $V_s$ misfit values for the best fit models (indicated by black stars in Figure 9b), supporting the advantage of incorporating component-consistent modal behavior in the inversion framework.

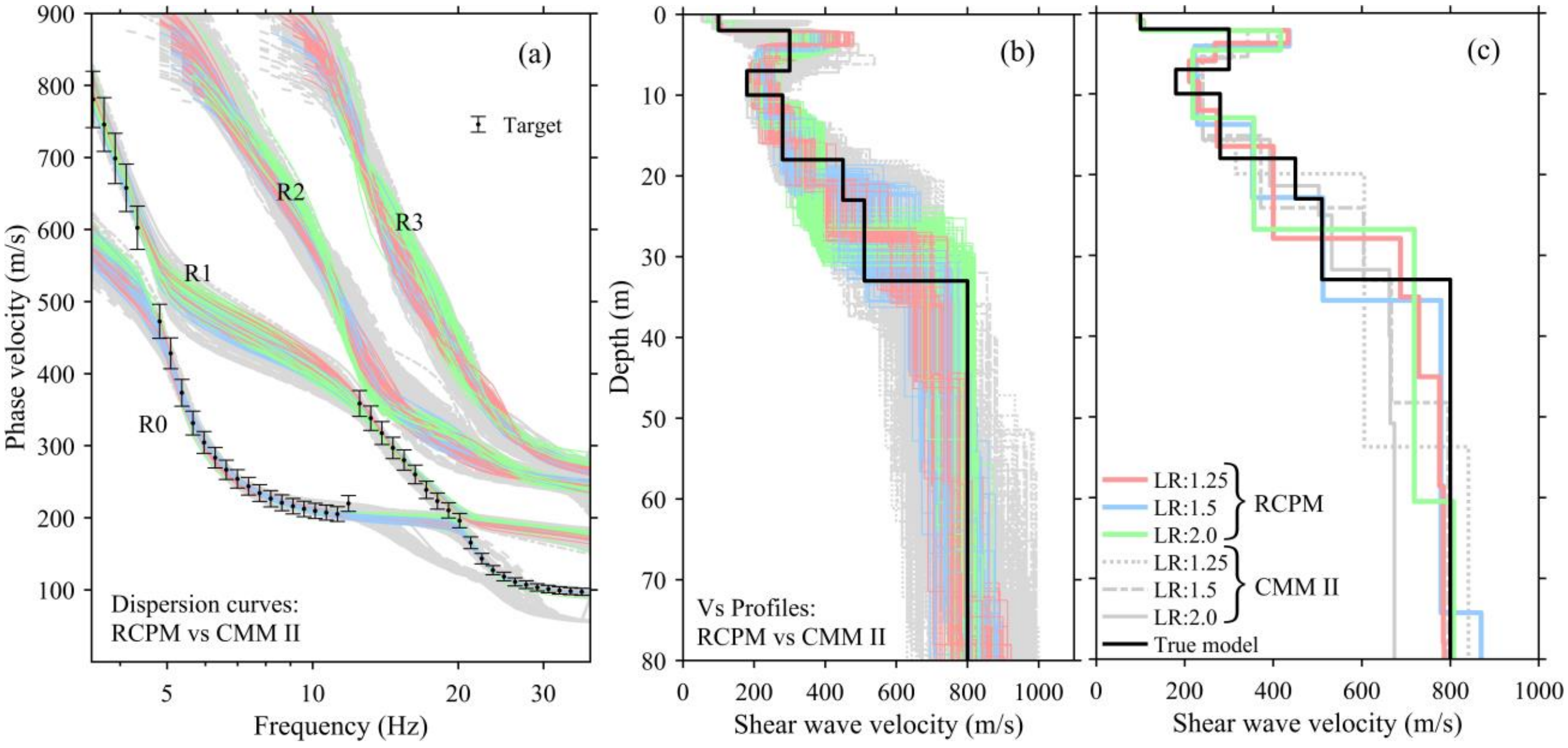

Figure 8. Inversion results for Ground Model III, using RCPM and CMM Scenario II: (a) target dispersion data (black points with uncertainty bars) and derived theoretical dispersion curves (R0-R3) corresponding to the 100 best-fit inverted profiles per LR (ten per trial), (b) ensemble of inverted $V_s$ profiles for different layering ratios (LR = 1.25, 1.5, and 2.0), along with the true model, and (c) comparison of inverted best-fit $V_s$ profiles per LR.

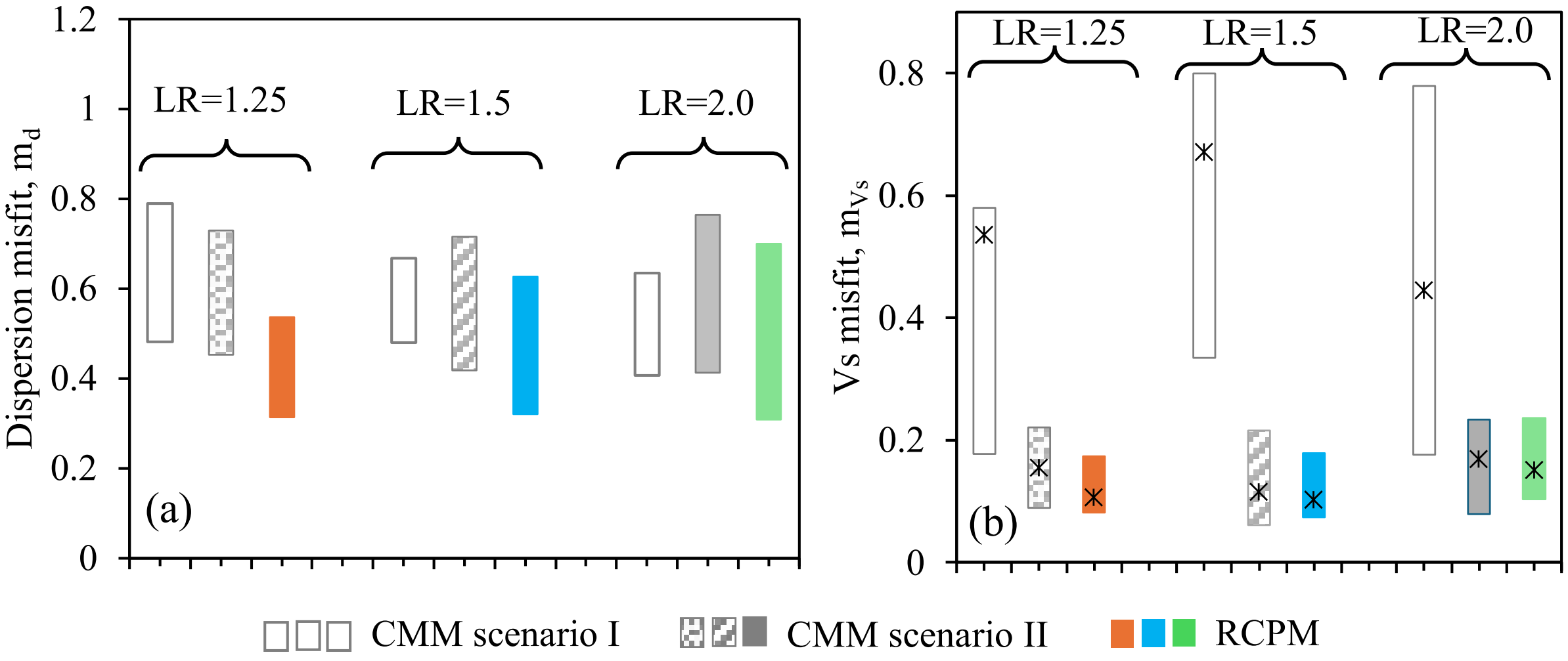

Figure 9. Comparison of RCPM and CMM inversion performance for Ground Model III: (a) ranges in dispersion misfit ($m_d$), and (b) ranges in shear wave velocity misfit ($m_{V_s}$) obtained from the 100 best-fit inverted profiles per LR (ten per trial). The bar ranges represent the minimum and maximum values across trials, with black asterisk markers in the $m_{V_s}$ bars indicating the lowest dispersion misfit profiles.

### 5.4. Discussion of RCPM Inversion Results for the Three Synthetic Ground Models

It has been shown that the modal energy distribution of Rayleigh waves may differ between vertical and radial components and is strongly dependent on subsurface stratigraphy. As a result, the appropriate modal representation cannot be reliably judged in advance, which makes conventional assumptions based on vertical-component behavior potentially inconsistent when applied to radial-component geophone or DAS-based measurements. The detailed investigation of three different synthetic ground models highlights that a component-consistent RCPM inversion framework provides more reliable results compared to conventional multi-modal inversion approaches when inverting radial-component dispersion data. In each case, the lowest dispersion misfit $V_s$ profiles from the RCPM approach were more accurate than those from the CMM approach. In CMM inversion, ambiguity in modal indexing can arise when modal transitions are not clearly identified in the experimental dispersion data. While sharp mode jumps can sometimes be interpreted easily, more complex scenarios, such as in Ground Model II, demonstrate that modal transitions may not follow a clear sequential pattern, with energy shifting between non-adjacent modes. In such cases, manual mode indexing becomes subjective and error prone. Furthermore, allowing the CMM inversion algorithm to automatically assign modes based solely on minimizing dispersion misfit can lead to ambiguous solutions, as observed in Ground Model III, where different modal interpretations yield similar dispersion misfit values. Therefore, a key limitation of CMM inversion is that it relies primarily on dispersion misfit, without explicitly accounting for modal energy distribution. Consequently, models that provide a good fit to the dispersion data may not correspond to physically consistent subsurface structures. This is reflected in the comparison between dispersion misfit and $V_s$ misfit, where profiles with minimum dispersion misfit do not necessarily yield the lowest $V_s$ misfit. Such discrepancies arise due to the uncertainty in dispersion measurements, which are simulated here through numerical modeling and assigned uncertainty bounds, mimicking real-world data conditions. In contrast, the proposed RCPM framework incorporates modal energy implicitly, enabling selection of dispersion branches that are physically meaningful rather than solely dispersion misfit driven. This reduces ambiguity in mode selection and improves consistency between data fit and subsurface model accuracy. Overall, the synthetic analyses demonstrate that the RCPM inversion provides a more robust and reliable approach for in-line radial measurements.

## 6. Application to Real Field DAS Data Sets

Following the successful RCPM validation using synthetic ground models, the RCPM inversion framework was applied to two field DAS datasets: Site A and Site B, as described below. The results are evaluated against available geologic information and invasive test results at each site to assess the reliability of the inverted subsurface $V_s$ profiles.

### 6.1. Site A: Hornsby Bend in Austin, Texas, USA

Site A is located at the Hornsby Bend test site in Austin, Texas, USA. DAS data was collected at this site during October, 2021 (Yust et al. 2022). A 200-m long fiber-optic cable manufactured by NanZee was installed in a shallow trench and backfilled. The data were recorded using an OptaSense ODH4+ interrogator unit, conFigured with a gauge length of 2.04 m and a channel spacing of 1.02 m. Based on tap-test results, 196 usable channels were identified in the DAS array. The surface waves were generated using the Thumper vibroseis mobile shaker truck from NHERI@UTexas (Stokoe et al. 2020) experimental facility. The source was operated in vertical mode with a linear frequency sweep from 5 to 200 Hz over a duration of 12 s. A total of 32 shot locations were used along the array, starting at −24 m off the end of the DAS array, with a constant spacing of 8 m. Further details of the acquisition procedure are provided in Yust et al (2024). For the present study, a subset of the DAS dataset consisting of the first 95 channels was selected. Six shot locations at −24, −16, −8, and 104, 112, and 120 m were used, providing approximately symmetric source coverage on both sides of the DAS array (Figure 10a). In addition to the surface-wave measurements, invasive data from cone penetration testing (CPT) and geotechnical borings are available at the site. CPT tests were conducted at 25-m intervals along the DAS array to an average depth of approximately 9.15 m. Two boreholes were also drilled along the DAS array at the 12.5 m and 137.5 m locations, both extending to a depth of around 23 m.

An example waterfall plot of the DAS waveforms recorded during the Thumper source sweep at −8 m and the corresponding dispersion image extracted from the waveforms using FDBF processing are shown in Figures 10b and 10c, respectively. The dispersion image shows a clear trend, likely representing R0 up to about 25 Hz, and a higher mode trend from about 25 to 40 Hz. The peak dispersion points extracted from the six selected shots are merged together to construct a statistical dispersion target, accounting for measurement uncertainty. The observed radial-component dispersion data and statistical dispersion target with error bars are shown in Figure 11a as gray circles and black markers, respectively. The proposed RCPM inversion was carried out using three different layering parameterizations, defined by LRs of 1.2, 1.3 and 1.5, with five independent trials conducted for each LR. An LR value of 2.0 was not considered in this case, as higher LR values result in fewer numbers of layers and, therefore, are more suitable for inverting deeper profiles. The maximum wavelength measured in this case was approximately 38 m; accordingly, the top of the half-space is constrained at 19 m. This depth range can be adequately represented using the selected lower LR values.

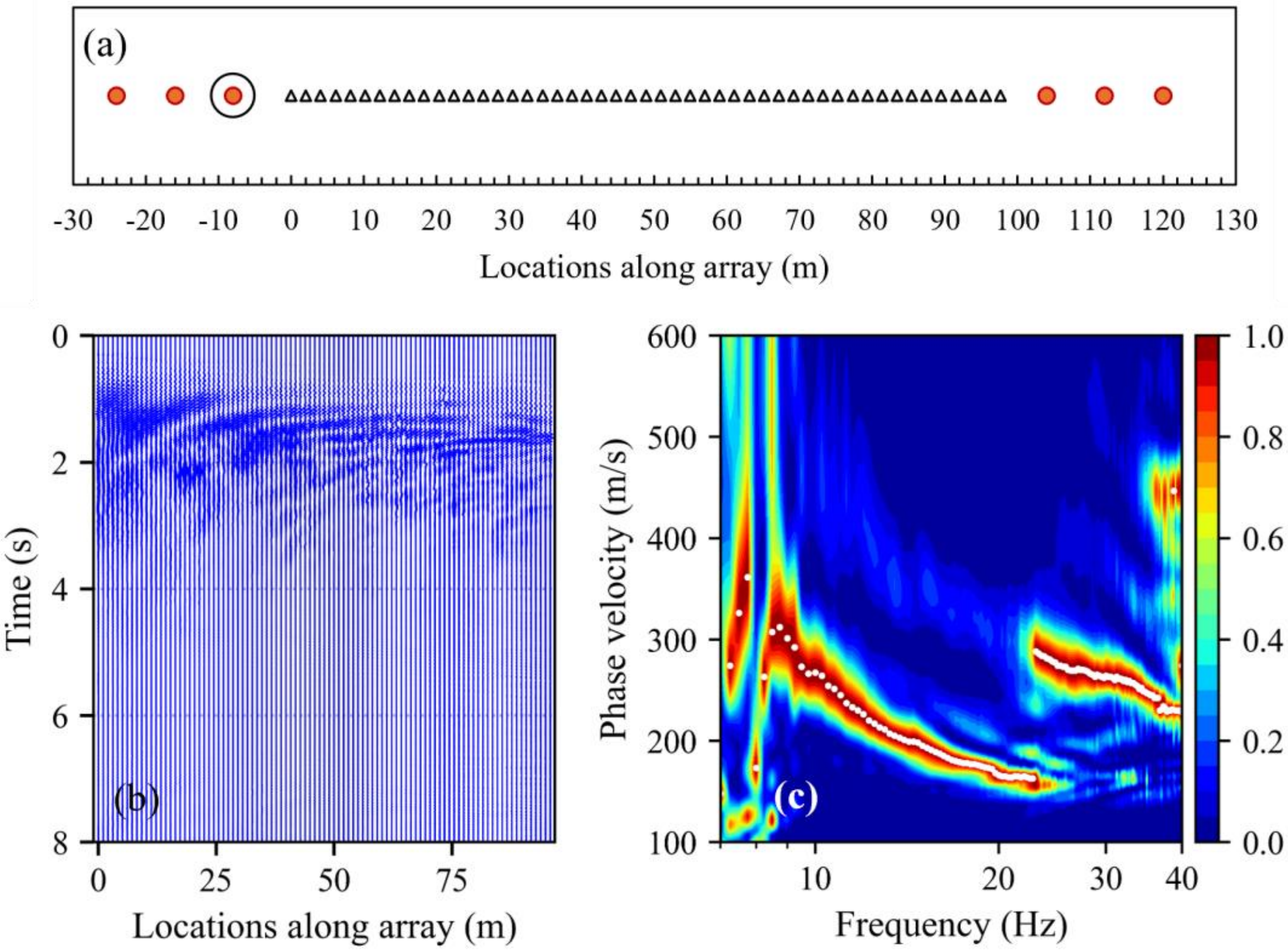

Figure 10. Example 95-channel DAS subarray and corresponding data used for RCPM inversions at Site A: (a) layout of the DAS array showing the selected channels (triangle symbols) and six shot locations (circular symbols) used to construct the inversion target, (b) recorded waveforms from a representative source sweep at −8 m, and (c) Rayleigh-wave radial component dispersion image extracted from the waveforms.

The single lowest misfit models for each trial of a given LR parameterization (five models per LR) were retained as possible representations of the subsurface beneath the DAS array. The theoretical dispersion curves from each of these models (Figure 11a) indicate that the higher-frequency experimental dispersion branch aligns more closely with the R2 mode. This further highlights the potential ambiguity in modal interpretation that can occur when using conventional approaches, wherein an analyst would have most likely assumed the higher mode experimental data was R1. The range in dispersion misfit values from the five lowest misfit models for each layering parameterization are presented in the legend below Figure 11a, and the corresponding inverted $V_s$ profiles are shown in Figure 11b. The inversion results consistently indicate a thin, soft near-surface layer, underlain by several layers of increasing stiffness. Across all parameterizations, the inverted profiles consistently identify a high $V_s$ layer interface at a depth of approximately 13.5–15 m. For validation of these results, $V_s$ profiles derived from correlations to CPT data and obtained from Yust et al. (2024) are plotted in Figure 11b. The comparison shows good agreement within the upper ~10 m, below which the CPT data does not extend. Hence, validation of the deeper RCPM inverted $V_s$ profiles can only be achieved by comparing the layering with that indicated in the closest borehole. The nearest borehole log at location 12.5 m indicates a sandy silty clay (CL–ML) layer extending to approximately 8.5 m depth, followed by clayey sand (SC) and dense SC layers, which overlay a shale layer located at approximately 13.4 m depth. The strong impedance contrast identified in the inverted $V_S$ profiles at about 13.5–15 m is in good agreement with the location of this shale layer, further supporting the reliability of the RCPM inversion.

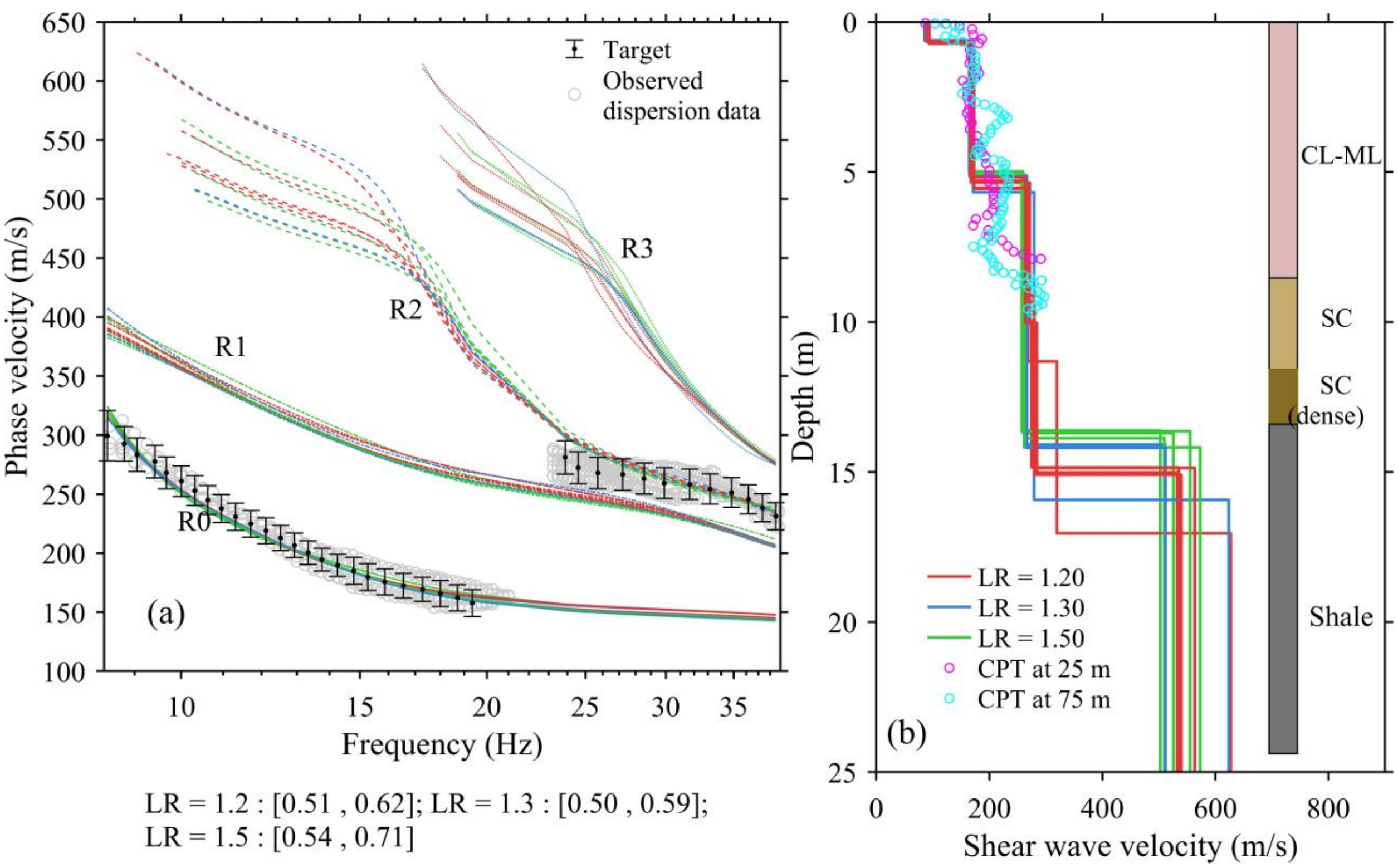

Figure 11. Inversion results for Site A using the RCPM approach: (a) theoretical dispersion curves corresponding to the inverted models for LR = 1.2, 1.3, and 1.5 parameterizations, plotted together with observed DAS dispersion data (gray circles) and inversion target with uncertainty bars, and (b) inverted shear wave velocity profiles for different LR parameterizations, compared with available CPT-derived Vs profiles at 25 m and 75 m along the DAS array and the interpreted lithology from a borehole near 12.5 m along the array.

### 6.2. Site B: Birds Landing - Sacramento, USA

Site B is located along the Sacramento River on Sherman Island in Sacramento, California, USA. The DAS data was collected during March 2025 as part of an experiment to monitor dynamic strains induced along a levee from the controlled explosive collapse of an electrical transmission tower (Panthi and Cox 2026). The site is predominantly composed of lean and organic clay in the near surface, underlain by a thick fibrous peat deposit, followed by delta mud deposits at greater depth. Accordingly, a thick low-velocity layer is expected beneath the surface. Approximately 1.4 km of fiber-optic cable was installed in a straight line parallel to the West Sherman Island Road levee. The DAS fiber-optic cable was a single-mode cable manufactured by TiniFiber, and was installed in a shallow trench of approximately 150 mm depth and subsequently backfilled and compacted to ensure adequate ground coupling. The cable was connected to an OptaSense ODH4+ interrogator unit. A gauge length of 2.04 m and a channel spacing of 1.02 m were used, resulting in a total of 1,435 channels. Active-source measurements were acquired at 10-m intervals along the cable using a 4.5-kg sledgehammer as a vertical source. Additionally, the surface waves generated from the collapse of the transmission tower, located approximately 200 m from the end of the DAS array, were also recorded.

For this study, active-source waveforms from a 48-channel subarray located between approximately 1230 m to 1280 m along the fiber were selected for dispersion processing. DAS records from six active shots (three on each side of the array) were used to construct the active-source dataset (Figure 12a). An example waterfall plot of the DAS waveforms from a sledgehammer shot located at -10 m and its corresponding dispersion image obtained from FDBF processing are shown in Figures 12b and 12c, respectively. The waterfall plot indicates that the wavefield requires more than 2 s to travel a distance of 50 m, reflecting the presence of low-velocity surface-waves. Consistent with this observation, the dispersion image exhibits phase velocities as low as approximately 20 m/s and frequencies as low as 1.5 Hz, along with evident mode jumps at higher frequencies. Although the presence of such low-frequency content is uncommon for a sledgehammer source, the corresponding maximum wavelengths (i.e., 15 – 20 m) are common for sledgehammer sources. Notably, a distinct discontinuity in the dispersion trend is observed around 1.5 Hz. To ensure a consistent dispersion trend at low frequencies, dispersion data derived from the tower implosion source was incorporated in the dispersion target. Specifically, data from array segments of 100 m and 200 m lengths, centered at the same subarray location, were utilized. Although no borehole data is available right at the DAS measurement location, a borehole log near the tower (extending to a depth of ~40 m) provides a useful basis for validation. Given the deltaic deposit, similar subsurface stratigraphy can reasonably be expected at the array location.

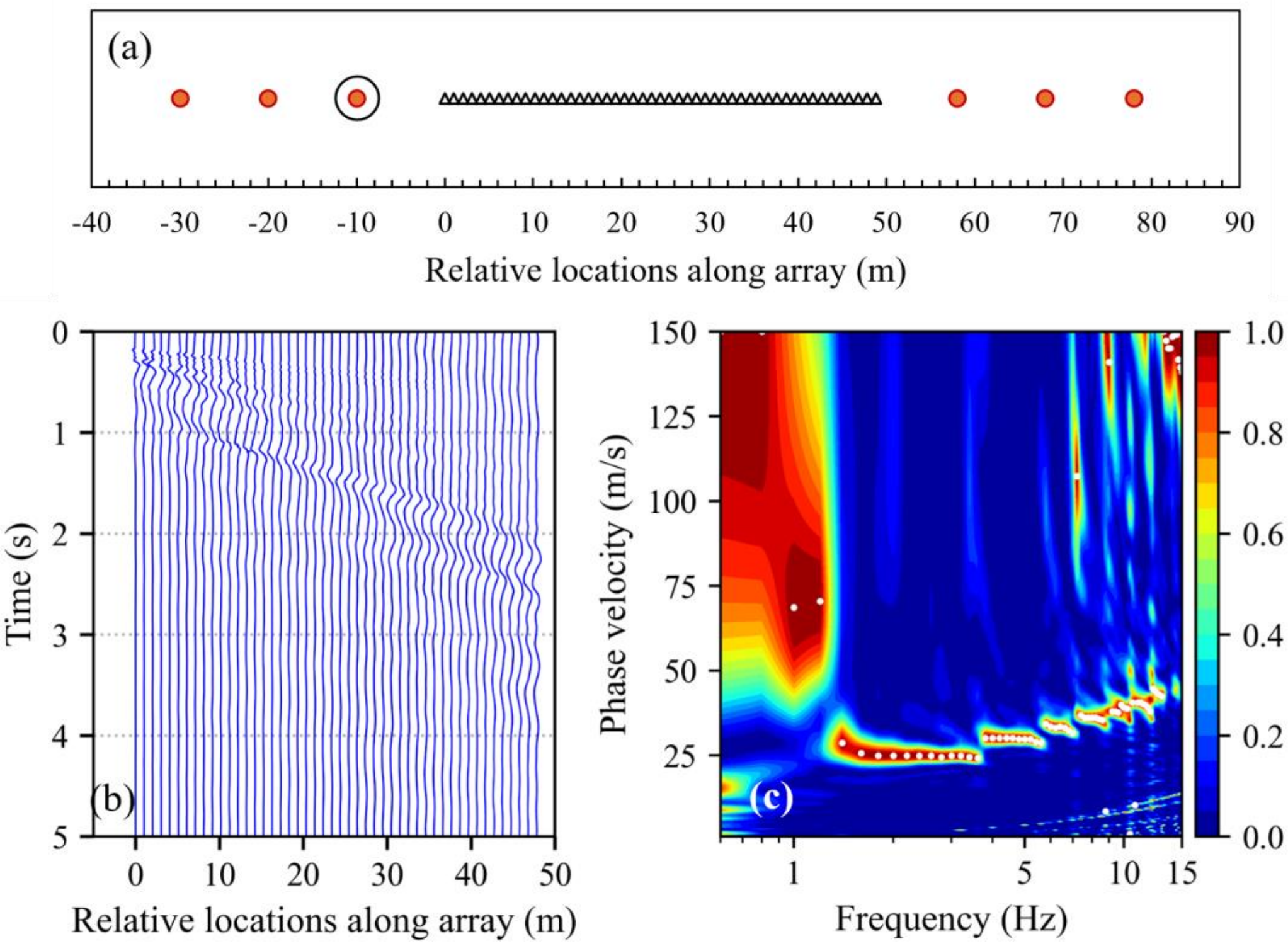

Figure 12. Example 48-channel DAS subarray used to extract active-source radial-component waveforms for dispersion processing and RCPM inversions at Site B: (a) layout of the DAS subarray and six shot locations used to construct the inversion target, (b) recorded waveforms from a representative sledgehammer shot at −10 m, and (c) corresponding Rayleigh-wave dispersion image.

The combined radial component DAS dispersion dataset, derived from both active sledgehammer shots and the tower-source, is presented in Figure 13a. In this Figure, gray circles denote all extracted dispersion data points, while black markers with error bars represent the statistical inversion target. The RCPM inversion was performed using three different layering parameterizations, defined by LRs of 1.25, 1.5, and 2.0, with five independent trials for each LR. The single lowest misfit models for each trial of a given LR parameterization (five models per LR) were retained as possible representations of the subsurface beneath the DAS array. The theoretical dispersion curves in Figure 13a indicate that, despite an apparent discontinuity in the target dispersion data near ~1.5 Hz, the inversion consistently interprets all dispersion data points below ~4 Hz as R0. This highlights a potential ambiguity in mode indexing when using a conventional multi-modal approach, where the dispersion data below ~1.5 Hz could potentially be incorrectly interpreted as R1 by the inversion analyst. In contrast, the proposed RCPM approach inherently resolves this ambiguity by identifying the dominating mode based on modal energy distribution. Additionally, it also consistently fits the higher-frequency portion of the target dataset by naturally incorporating higher modes up to R4 without requiring manual modal indexing.

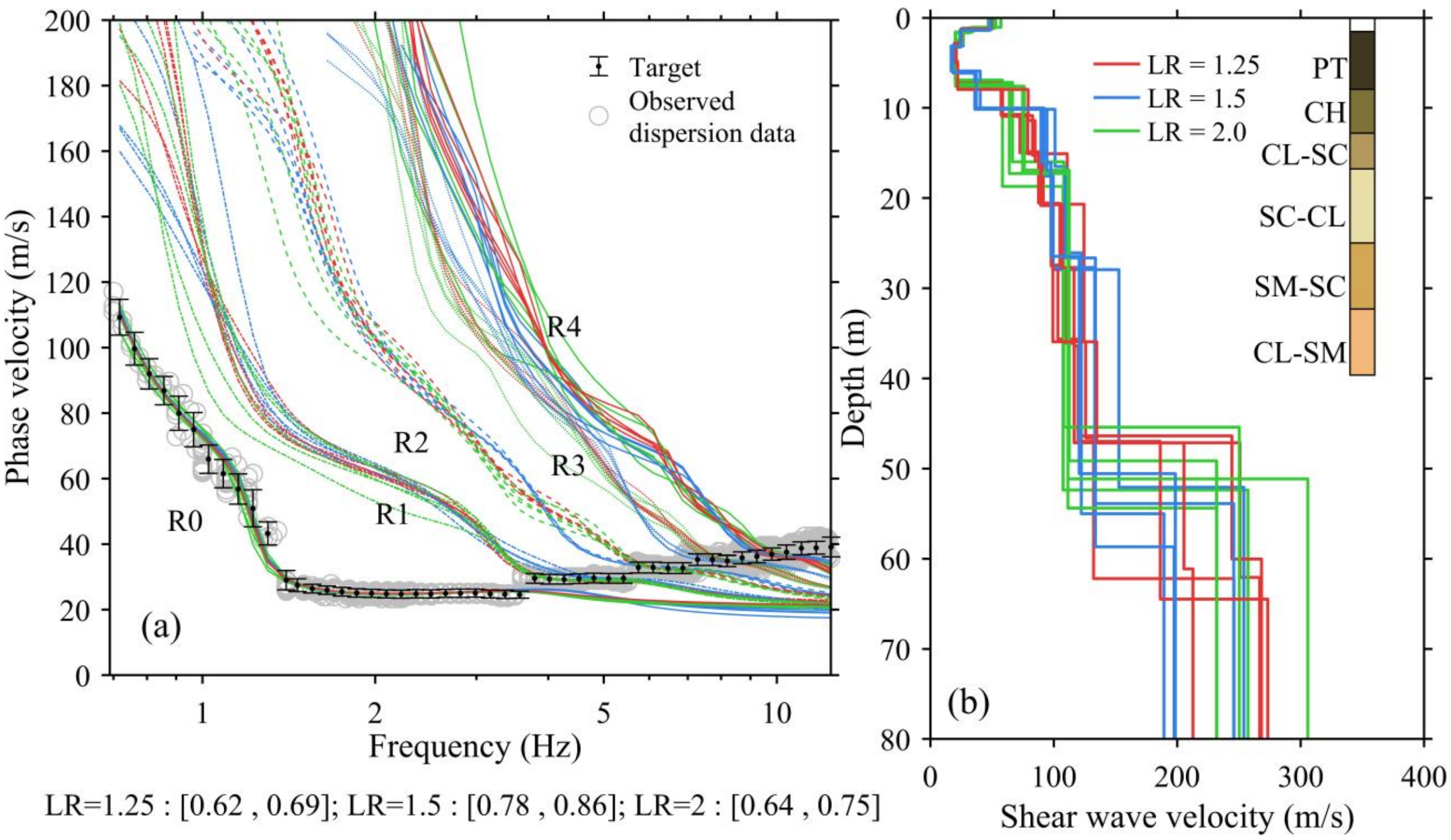

Figure 13. Inversion results for Site B using the RCPM approach: (a) theoretical dispersion curves corresponding to the inverted models for LR = 1.25, 1.5, and 2.0 parameterizations, plotted together with observed DAS dispersion data (gray circles) and inversion target with uncertainty bars, and (b) inverted shear wave velocity profiles for different LR parameterizations, compared with the interpreted lithology from a borehole near the tower.

The corresponding inverted $V_s$ profiles in Figures 13b consistently indicate a stiffer near-surface layer, followed by a thick, very-low velocity layer extending to approximately 8 m depth, with $V_s$ as low as ~18 m/s. The borehole log is in good agreement with the $V_s$ profiles, indicating sandy lean clay mixed with peat in the top 1.5–2 m, followed by an approximately 6.5-m thick layer of soft organic peat (PT), extending to a depth of 8 m. While $V_s$ ~ 18 m/s is exceptionally low, it is consistent with values for extremely soft, compressible soils, such as peat (Trafford and Long 2020). Below this PT layer, the borehole log indicates layers that more or less consistently transition from high-plasticity/fat clay (CH) to coarser-grain materials like lean clay mixed with clayey sand (CL-SC), to clayey sand mixed with lean clay (SC-CL), to silty sand mixed with clayey sand (SM-SC). Over these same depth intervals, the $V_s$ increases gradually with depth as the soil gradually coarsens. While the inverted Vs profiles indicate stiffer materials starting around 45 m, the borehole log does not extend this deep. The inverted Vs profiles are also consistent across different parameterizations (LR = 1.25, 1.5, and 2.0), with greater uncertainty in the depth range of approximately 45–60 m, which is associated with reduced resolution at greater depths. Overall, the consistency between the inverted $V_s$ profiles and the borehole data provide some verification for the reliability of the proposed RCPM inversion framework in capturing the key stratigraphic features, although it would have been nice if invasive $V_s$ measurements were available for complete validation.

# 7. Conclusion

DAS has emerged as a powerful tool for near-surface site characterization, enabling dense spatial sampling of wavefields over large distances. In typical configurations used for active-source surface wave testing, DAS waveform measurements predominantly capture the axial strain response, which is governed by the radial component of Rayleigh wave motion. However, when inverting radial-component DAS dispersion data analysts presently rely on conventional inversion frameworks, such as fundamental-mode or multi-modal inversions that are more consistent with vertical-component dispersion data and rely on explicit and often subjective mode identification. This study demonstrated that such conventional approaches may face significant challenges when applied to radial-component data, primarily due to component-dependent modal behavior, modal osculation, and ambiguity in mode identification under complex subsurface conditions. Consequently, incorrect modal interpretation in practical field applications can lead to physically inconsistent subsurface velocity profiles, thereby limiting the true potential of high-resolution DAS measurements.

To address these limitations, this study presented a radial-component predominant-mode (RCPM) inversion framework specifically designed for DAS-based surface-wave measurements that explicitly accounts for the sensitivity of the Rayleigh-wave radial component under a vertical-source configuration. The proposed framework incorporates radial-component modal energy characteristics directly into the inversion process through a forward-modeling approach in which the predominant mode at each frequency is inherently identified based on the highest surface modal participation amplitude and matched with the dominant trends observed in the measured dispersion data. As a result, the framework eliminates the need for explicit modal indexing, provides a more physically consistent interpretation of measured radial-component dispersion characteristics, and substantially reduces reliance on subjective analyst-driven decisions. Systematic evaluation of the RCPM approach using three synthetic ground models highlights that modal energy distribution varies between vertical and radial components under a vertical load and is strongly influenced by subsurface stratigraphy. These differences can lead to mode indexing ambiguity in conventional multimodal (CMM) inversion, particularly in the presence of strong velocity contrast, low-velocity layers, and high-velocity layers. This study further demonstrated that modal transitions in the simulated radial response may not follow a clear sequential pattern, with energy shifting between non-adjacent modes. Under such conditions, manual mode indexing required by CMM inversions becomes subjective and error prone. Furthermore, allowing CMM inversion algorithms to assign modes solely through dispersion-misfit minimization may lead to ambiguous solutions, where different modal interpretations produce similar dispersion misfit values. Consequently, CMM inversion results that provide a good fit to the dispersion data may not correspond to physically consistent subsurface profiles. In contrast, the proposed RCPM framework consistently captures the correct modal behavior across all tested scenarios without requiring manual mode indexing. The inversion results not only achieve comparable or lower dispersion misfit, but also yield $V_S$ profiles that are more physically consistent and stable across different parameterizations. Application to two field DAS datasets further demonstrates the robustness and practical applicability of the proposed RCPM inversion method. The RCPM inversion results successfully capture key subsurface features, including soft near-surface layers and deeper velocity contrasts, and show good agreement with available invasive measurements, such as CPT and borehole data. Importantly, the framework effectively handles complex dispersion characteristics, including apparent discontinuities, low-frequency ambiguities, and higher-mode dominance, without requiring subjective interpretation or manual mode selection.

Overall, the findings emphasize the importance of adopting component-consistent forward modeling and inversion strategies for DAS-based surface-wave analysis. By explicitly accounting for the physics of radial-component measurements, the proposed RCPM framework offers a fundamentally improved approach for interpreting dispersion data extracted from DAS waveform measurements. It reduces ambiguity in modal identification, enhances inversion stability, and provides more reliable estimates of subsurface $V_S$ profiles. Thus, the proposed RCPM inversion approach has significant implications for near-surface geophysics and geotechnical site characterization using DAS.